\documentclass[12pt]{article}
\usepackage{amssymb, amsmath, amsthm,verbatim}
\usepackage{lscape} 
\usepackage{mathdots}
\usepackage{multirow}
\usepackage{thmtools}
\usepackage{mathtools}
\usepackage[makeroom]{cancel}
\usepackage{appendix}
\usepackage{bm}
\usepackage{bbm}
\usepackage{caption}
\usepackage{subcaption}

\DeclareCaptionType{intervention}
\usepackage{enumerate} 
\usepackage{graphicx}
\usepackage{natbib}
\usepackage{setspace}
\usepackage[a-1b]{pdfx} 
\usepackage{hyperref} 
\usepackage{xcolor}
\usepackage{array}
\usepackage{rotating}
\hypersetup{
    colorlinks,
    linkcolor={red!50!black},
    citecolor={blue!50!black},
    urlcolor={blue!80!black}
}

\usepackage{tocloft}
\usepackage[T1]{fontenc}
\usepackage[utf8]{inputenc}

\usepackage[explicit]{titlesec}
\usepackage{tikz}
\usetikzlibrary{bending}
\usetikzlibrary{shapes,snakes}
\usetikzlibrary{external}
\usepackage{mathrsfs}

\usepackage[top=1.2in, bottom=1.2in, left=1.2in, right=1.2in]{geometry}
\pdfoutput=1

\def\phi{\varphi}

\def\bar{\overline}

\newtheorem{theorem}{Theorem}
\newtheorem{lemma}{Lemma}
\newtheorem{proposition}{Proposition}

\newtheorem{claim}{Claim}

\theoremstyle{definition}
\newtheorem{example}{Example}
\theoremstyle{plain} 

\begin{document}
\title{Misspecified Model Estimation and Its Impact on Predictions  }
\author{Junnan He\footnote{Department of Economics, Sciences Po, junnan.he@sciencespo.fr.}
\and Lin Hu\footnote{Research School of Finance, Actuarial Studies and Statistics, Australian National University, 
\href{mailto:lin.hu@anu.edu.au}{\color{black}lin.hu@anu.edu.au.}}
\and Matthew Kovach\footnote{Department of Economics, Purdue University, \href{mailto:mlkovach@purdue.edu}{\color{black}mlkovach@purdue.edu}.} 
\and Anqi Li\footnote{Department of Economics,  University of Waterloo, \href{mailto:angellianqi@gmail.com}{\color{black}angellianqi@gmail.com}. }
}
\date{ }

\maketitle
\thispagestyle{empty}
\begin{abstract}
We study a linear statistical model where outcomes depend on regressors with fixed population coefficients and observation-specific latent coefficients, along with measurement errors. A decision-maker estimates population coefficients and uses the estimates to predict the latent coefficients for a given observation. We analyze how misspecification of some population coefficients distorts predictions, investigating comparative statics with respect to:  (1) residual information in regressors associated with misspecified coefficients after projecting out those associated with free coefficients, (2) alignment between misspecification vector and latent-to-coefficient mapping. Applications include employee rating with unconscious bias and LLM-mediated consumer research.


\end{abstract}

\newpage
\setcounter{page}{1}

\section{Introduction}\label{sec_intro}
This paper analyzes a linear statistical model where outcomes depend on regressors with fixed population coefficients and observation-specific latent coefficients, along with measurement errors. A decision-maker (DM) estimates population coefficients and uses the estimates to predict the latent coefficients for a given observation. During estimation, the DM misspecifies some population coefficients by fixing their values to those differ from the truth.  We analyze how such misspecification distorts the estimation of the remaining population coefficients and the prediction of latent coefficients.

Our estimation-prediction paradigm is common in reality. Consider an HR department of a company where raters of varying seniority evaluate a large population of minority workers. A rater's evaluation of a worker equals the sum of the rater's bias, the worker's productivity, and a rating error. The manager of the HR department observes ratings but not productivity. Having long prior interactions with senior raters, the manager feels confident in his knowledge of their biases, even though this belief is not grounded in truth. For instance, the manager may exhibit an implicit bias against minorities when conducting evaluations himself, or he may mistakenly view prejudiced senior raters as neutral, particularly in complex hierarchies that lack accountability and transparency. By contrast, the manager is unfamiliar with the junior raters and must estimate their biases. Based on these estimates, he predicts each worker’s productivity and determines compensation. A systemic, persistent compensation distortion arises if misspecification of senior raters' biases distorts prediction of minority workers' productivity.

In settings such as the above, we compare the prediction of the latent coefficients under misspecification with the one obtained in its absence. We fully characterize the prediction distortion and investigate comprehensive comparative statics. Our approach applies regardless of dimensionality and facilitates normative analysis.

Our DM’s worldview is distorted by the misspecification constraint. Given that worldview, he first conducts constrained generalized least squares (GLS) estimation of the population coefficients and then the best linear prediction (BLP) of the latent coefficients. Although these procedures depend only on the first two moments of the environment, in Gaussian settings they are equivalent to the Kullback-Leibler divergence minimizer and the maximum a posteriori estimator, respectively. The two stages can also be compressed into one, yielding the best linear unbiased predictor in mixed models and, under Gaussian assumptions, the maximizer of a joint penalized likelihood. Our framework therefore admits multiple interpretations, ranging from classical linear estimation and prediction to misspecified Bayesian learning and machine learning. Importantly, from the DM’s viewpoint, both the estimator and the predictor are optimal.

We adopt the GLS-BLP interpretation, formalizing each stages as minimizing a quadratic loss and studying the structural link between them. To begin, note that GLS estimation minimizes the squared outcome mean, weighted by the outcome covariance matrix. In an infinite population, the estimated outcome mean is the residual from projecting the regressors associated with the misspecified coefficients onto the column space spanned by those associated with the free coefficients. In the second stage, this residual is projected onto the latent space to form the BLP. Crucially, both stages use the outcome covariance matrix as the weighting matrix and therefore employ the same linear projector. Moreover, we can decompose the latent space into two subspaces: one spanned by the misspecified-coefficient regressors and the other by the free-coefficient regressors. The projection of the first-stage residual onto the free-regressor space is zero, while that onto the misspecified-regressor space yields a quadratic form—specifically, the Schur complement representing the residual information in the misspecified regressors after projecting out the component in the free-regressor space.

The overall prediction distortion admits a simple formula: the vector of misspecified population coefficients is mapped onto the latent space via a projection filter. A key component of this filter is the Schur complement, which depends exclusively on the regressors and the error precision matrix. Another component is a regularizer arising from the mapping from the population-coefficient space to the latent space. We thus isolate the role of each model ingredient in the prediction distortion. Our comparative static analysis varies (1) the Schur complement and (2) the alignment between the misspecification vector and the population-to-latent space mapping.

When the DM misspecifies multiple population coefficients, both concepts above become matrix-valued objects, making the derivation of monotone comparative statics challenging. We note that multidimensional misspecification is common in practice; in our leading example, the manager may misspecify the biases of multiple senior raters. We develop approaches to overcome the challenges it poses.

For comparative statics regarding the Schur complement, we consider adding free information to the environment, including adding free regressors, adding observations, introducing novel information sources, and enhancing error precision. For each change, we identify its impact on the Schur complement, focusing on conditions that yield a null impact. In practice, these null conditions often hold for policy interventions or technological changes viewed as progress in isolation. 
 For instance, adding new regressors that are linearly independent of existing ones typically enhances the explanatory power of standard statistical models. In our setting, however, it has no impact on the Schur complement and, consequently, on the prediction distortion. Likewise, introducing novel information sources has no impact, provided the errors associated with existing and new sources are independent. Finally, adding observations and enhancing error precision both increase the Schur complement in the Loewner order and therefore represent informational improvements. Under multidimensional misspecification, however, their impacts on prediction distortion are generally ambiguous—a result that carries potentially counterintuitive policy implications.

The above ambiguity can be traced to the misalignment between the misspecification vector and the population-to-latent space mapping. We develop our notion of alignment by projecting the misspecification vector onto the eigenbasis of the regularizer. Under this formulation, rotations of the misspecification vector—reflecting realignment with the regularizer’s eigenvectors—carry clear, interpretable consequences. We examine the worst-case impact of such a rotation---measured by the semi-elasticity of the prediction distortion's norm. We find that when there are more misspecified coefficients than the latent space's dimensionality, the worst-case elasticity is unbounded. Conversely, when this dimensionality condition fails, the worst-case elasticity may be uniformly bounded.


Our results speak to discrimination arising from unconscious biases. When raters are biased against minorities in reality but are perceived as neutral, the manager’s prediction of minority productivity is downward distorted relative to the truth or to its majority counterpart.   Policies aimed at addressing discrimination---such as debiasing raters or introducing new, unfamiliar raters---may prove ineffective in mitigating this distortion. Informational improvements that result from more frequent performance evaluations or precision-enhancing rater training may even prove counterproductive. Since these changes are costly to implement, our results identify   potentially wasteful interventions and provide a framework for cost-benefit analysis. Finally, training aimed at recalibrating the manager’s beliefs may cause rotations in the misspecification vector. The worst-case impact is unbounded if the number of misspecified senior raters exceeds the dimensionality of worker skills.



Our results also inform the design of consumer research mediated by Large Language Models (LLMs). A downstream company conducts research on consumer preferences by querying an LLM. The regressor matrix captures prompts or, more generally, the design of research questionnaires. Inquiry outcomes for an individual consumer depend on these prompts, their population effects, and the consumer's latent preference. Among all population coefficieints, some are fine-tuned by upstream foundation model developers to achieve alignment objectives, and recovering them is challenging for downstream actors due to their high dimensionality and opacity. We thus model the downstream company as misspecifying the fine-tuned population coefficients while estimating those that are unique to its own research design. We examine how such misspecification distorts the prediction of consumer preferences and subsequent activities such as pricing and targeting.

Our analysis provides conditions for a non-vanishing prediction distortion—a form of the ``alignment tax'' discussed in the AI safety literature. We identify several developments—such as prompt engineering, the use of synthetic data, in-house research integration, and upstream anti-hallucination measures—that may be ineffective or even aggravate this tax. In our framework, these correspond to adding free regressors, adding observations, introducing novel information sources, and enhancing error precision. 
Furthermore, adjustments to upstream alignment objectives can cause rotations in the fine-tuned coefficient vector, which carry an unbounded worst-case impact. This result follows from the astronomical dimensionality of fine-tuning relative to that of consumer preferences—a defining characteristic of the AI era.

 Section \ref{sec_literature} reviews the related literature. The Online Appendix performs robustness checks and presents a further application of our framework to misperceived media bias.

\section{Model}\label{sec_model}
\subsection{Setup}\label{sec_setup}
Consider a statistical model 
\begin{equation}\label{eq:model}
Y_i = X\beta+Z\omega_i + \epsilon_i,
\end{equation}
where $i$ indexes observations and its population is taken to be infinity. $Y_i \in \mathbb{R}^T$ is the vector of outcome variables, and $X \in \mathbb{R}^{T\times s}$ is the matrix of regressors. 

There are two coefficient vectors: $\beta$ and $\omega_i$. $\beta \in \mathbb{R}^s$ collects the  population coefficients and is constant across observations; $\omega_i \in \mathbb{R}^r$ collects the latent coefficients for observation $i$ and is i.i.d. across observations.  With the standardization in Online Appendix \ref{sec_standardization},  one can normalize the true population coefficients $\beta^*$ to zero, and the distribution of latent coefficients to have mean zero and identity covariance $\Omega=Id_r$. 

The matrix $Z \in \mathbb{R}^{T\times r}$ maps latent coefficients to outcomes. The vector $\epsilon_i \in \mathbb{R}^T$ collects the measurement errors for observation $i$; it is i.i.d. across observations with mean zero and a positive-definite covariance matrix $\Sigma \in \mathbb{R}^{T \times T}$. Across population, $\omega_i$ and $\epsilon_j$ are independent for all $i$ and $j$. 

The rank conditions are $r\leq s \leq T$. The assumption that $r\leq s$ is without loss of generality (w.l.o.g), because one can always include $Z$ in the regressor matrix. Our analysis requires that $X$ has full column rank $s$, which holds only if $s \leq T$.   

A decision maker (DM) observes $X$ and $Y_i$'s, and he knows the model structure, including $Z$, $\Omega$, and $\Sigma$. The DM learns about $\beta$ and $\omega_i$ in two stages, by first estimating $\beta$ and then using this estimate to form a prediction of $\omega_i$, where $i$ indexes an arbitrary observation. We formulate both  stages as minimizing quadratic losses and provide interpretations of this formulation toward the end of this section. 

During the estimation stage, the DM misspecifies the first $q$ population coefficients $\beta_{m}\coloneqq \beta_{1:q}$ by fixing them at $\delta \neq 0$. The remaining coefficients $\beta_{f}\coloneqq \beta_{q+1:s}$ are estimated by solving
\begin{equation}\label{eq:gls}
\min_{\beta \in \mathbb{R}^s: \; \beta_{m}=\delta} \beta'X' M^{-1} X\beta,
\end{equation}
where $M \coloneqq \Sigma + ZZ'$ denotes the covariance matrix of $Y_i$  given $X\beta$. The the subscripts $m$ and $f$ stand for ``misspecified'' and ``free,'' respectively. Letting $\hat{\beta}_{f}$ denote the solution, the full coefficient estimate is $\hat{\beta} = [\delta^{\top}, \; \hat{\beta}_f^{\top}]^{\top}$. 

Given $\hat{\beta}$, the DM forms a prediction of $\omega_i$ by solving 
\begin{equation}\label{eq:blp}
\min_{\hat{\omega}_i \in \mathbb{R}^r} \ \hat{\omega}_i'\hat{\omega}_i + (Y_i - X\hat{\beta} - Z\hat{\omega}_i)'\Sigma^{-1}(Y_i - X\hat{\beta} - Z\hat{\omega}_i).
\end{equation}
The solution, as a random variable of $Y_i$, is denoted by $\hat{\omega}(Y_i)$.

We examine how coefficient misspecification as above distorts estimation and prediction. We compare our framework to a correctly specified benchmark in which $\beta_m$ is set equal to its true value 0, or one in which the constraint on $\beta_m$ is dropped from estimation. Under both benchmarks, the first-stage estimator coincides with the true population coefficients 0 in the infinite-population limit. The resulting second-stage predictor is denoted by $\omega^0(Y_i)$. The distortion in the predictor due to misspecification: 
\[
\Delta\hat{\omega}(Y_i) \coloneqq \hat{\omega}(Y_i)- \omega^{\,0}(Y_i),
\]
is the focus of the current study. Hereafter, it is simply referred to as ``the prediction distortion,'' or simply  ``the distortion.'' 

\subsection{Discussion}
Our framework encompasses several familiar estimation and prediction paradigms. Under assumptions about the first two moments alone,  \eqref{eq:gls} can be interpreted as constrained generalized least squares (GLS) estimation of $\beta$ under an infinite population, while \eqref{eq:blp} is the  best linear prediction (BLP) of $\omega_i$, minimizing mean-squared error.\footnote{
For a given sample size $N$,  it is well known that the best linear unbiased estimator under the constraints is the constrained GLS estimator,    which solves 
\[
\min_{\beta:\,\beta_m=\delta}
\sum_{i=1}^N (Y_i-X\beta)'M^{-1}(Y_i-X\beta).
\]
Under the normalization that \(\beta^*=0\), we have \(\mathbb{E}[Y_i]=0\), so the sample objective converges to~\eqref{eq:gls} as $N \rightarrow \infty$. Given $\hat{\beta}$, the BLP minimizes mean-squared error: 
\[
\min_{L_{\hat{\beta}}(Y_i)}\ \mathbb{E}\big[(\omega_i-L_{\hat{\beta}}(Y_i))'(\omega_i-L_{\hat{\beta}}(Y_i))\big],
\]
where \(L_{\hat{\beta}}(Y_i)\) is linear in the residual \(Y_i-X\hat{\beta}\). Operationally, this problem is equivalent to~\eqref{eq:blp}. 
}  
 While we adopt a two-stage interpretation, one can alternatively formulate the problem as a single joint minimization: 
\begin{equation}
\min_{\beta: \beta_m=\delta, L(Y_i) \in \mathcal{A}} \mathbb{E}\left[L(Y_i)'L(Y_i)+(Y_i-X\beta-ZL(Y_i))'\Sigma^{-1}(Y_i-X\beta-ZL(Y_i))\right],
\end{equation}
where $\mathcal{A}$ denotes the set of affine maps from the outcome space to the latent space. 
The solution, coinciding with that of \eqref{eq:gls} and \eqref{eq:blp}, is the best linear unbiased predictor (BLUP) in mixed models \citep{henderson1975best}.

Now suppose, in addition, that both the latent coefficients and the measurement errors are Gaussian. Then \eqref{eq:gls} is equivalent to constrained maximum likelihood estimation (MLE), or equivalently to Kullback-Leibler (KL) divergence minimization. \eqref{eq:blp} is maximum a posteriori (MAP) estimation, which maximizes the prior-regulated likelihood. These formulations are commonly seen in modern learning frameworks such as misspecified Bayesian learning, variational learning, and machine learning \citep{berk1966limiting, cinelli2021variational}.\footnote{Under the misspecified Bayesian learning interpretation, a DM 
 holds a prior over the outcome-generating distributions parameterized by $\beta$ that excludes the true distribution from its support. By \citet{berk1966limiting}, the posterior upon repeated draws from the true distribution concentrates on the parameters that minimize the KL divergence from the true distribution.
For Gaussian distributions with a known variance matrix \(M\), 
\[\mathrm{D_{KL}}\!\left(\mathcal{N}(0,M) \| \; \mathcal{N}(X\beta,M) \right)
=
\frac{1}{2}\,\beta'X'M^{-1}X\beta
\;+\; \text{terms that depend only on } M,
\]
so the KL minimizer solves~\eqref{eq:gls}. 
\label{fn_berk}} An equivalent grand problem is to maximize the joint penalized likelihood: 
\begin{equation}\label{eq:joint_program}
\min_{\beta \in \mathbb{R}^s: \beta_m=\delta, \;\{\hat{\omega}_i\}\;}
\sum_{i=1}
\left[
\hat{\omega}_i'\hat{\omega}_i
+
\bigl(Y_i - X\beta - Z\hat{\omega}_i\bigr)'\Sigma^{-1}\bigl(Y_i - X\beta - Z\hat{\omega}_i\bigr)
\right].
\end{equation}

In what follows, we main the GLS-BLP interpretation, since it relies only on moment assumptions and highlights the structural connection between the estimation of population coefficients and the prediction of latent coefficients. 

Our DM's worldview is pinned down by the misspecification constraint $\beta_m=\delta$. Conditional on this worldview, his estimator and predictor are optimal. 

The baseline setting makes several simplifications. Specifically, we hold the regressor matrix $X$ fixed across observations, represent coefficient misspecification as $\beta_m=\delta \neq 0$, and assume that the DM knows the error covariance $\Sigma$. In the Online Appendix, we relax these assumptions by considering observation-specific regressors and general linear misspecifications.  We also study an extension in which $\Sigma$ is jointly estimated  along with $\beta$.  

\subsection{Examples}\label{sec_example}
The following examples serve to illustrate our framework and results. 

\begin{example}[Employee rating with unconscious bias]\label{exm:labor}
Raters $j=1,\ldots,s$ conduct performance evaluations of minority workers indexed by $i$. 
Rater $j$'s evaluation of worker $i$ is \[Y_{ij}=\beta_j+\omega_i+\epsilon_{ij},\]
where $\beta_j\in\mathbb{R}$ denotes rater $j$'s bias, $\omega_i\in\mathbb{R}$ denotes worker $i$'s productivity, and $\epsilon_{ij} \in \mathbb{R}$ is a rating error that is independent across raters and workers, with mean zero and variance $\sigma_j^2$. A senior manager collects all raters' evaluations for worker $i$ into a vector $Y_i\in\mathbb{R}^s$. Then $Y_i$ follows the distribution in \eqref{eq:model} with
\[
X=Id_s,\quad Z=\mathbf{1}_s,\quad \text{and}\quad \Sigma=\operatorname{diag}(\sigma_1^2,\cdots,\sigma_s^2).
\]

Among all raters, the first $q$ are senior and the remaining are junior. Having long prior interactions with the senior raters, the manager feels confident that he knows their biases, although this belief is  is not grounded in truth. For instance, the manager may be subject to implicit bias when conducting employee evaluations himself  \citep{greenwald1998measuring}. Alternatively, the other senior raters may hold prejudice, stereotypes, and intergroup hostility against minorities \citep{becker2010economics, bordalo2016stereotypes, tajfel1982social}. Still,  the manager views them as neutral---a well-documented phenomenon in large organizations with complex hierarchies and limited transparency \citep{bohren2025systemic}. 

By contrast, the manager is unfamiliar with the junior raters and must estimate their biases. Based on these estimates, he predicts worker $i$'s productivity and determines compensation accordingly. Prediction errors generate a compensation distortion relative to the benchmark without misspecification, or to majority workers whose evaluations are often subject to a null or positive unconscious bias. 
\end{example}

\begin{example}[LLM-mediated consumer research]\label{exm:ai}
A consumer research company queries a large language model (LLM) about consumer preferences. Let $\omega_i \in \mathbb{R}^r$  denote consumer $i$'s preference, which may be multidimensional (on the order of $10^2$). The mapping from $\omega_i$ to the inquiry outcome $Y_i \in \mathbb{R}^T$ is given by $Z \in \mathbb{R}^{T\times r}$.

The matrix $X \in \mathbb{R}^{T\times s}$ represents prompts, or more broadly how inquiries are designed and encoded. The inquiry about each consumer $i$ consists of $T$ questions, each prompting on a  subset of the $s$ prompting directions. For instance, if
    \[
X=
\begin{bmatrix}
1 & 0 & 0 \\
0 & 1 & 1 \\
0 & 1 & 0 \\
1 & 0 & 1
\end{bmatrix},
\]
then the inquiry consists of four questions, the first prompting on dimension 1, the second prompting on dimensions 2 and 3, the third prompting on dimension 2, and the last prompting on dimensions 1 and 3. 

The population effects of prompting on inquiry outcomes is summarized by the vector $\beta \in \mathbb{R}^s$. Among these coefficients, the first $q$ are fine-tuned by upstream foundation model developers to satisfy alignment objectives, such as protecting consumer privacy or safety, or reducing sensitivity to emotionally charged or politically biased language \citep{bai2022constitutional, openai2024gpt4o, anthropic25}. The extent of fine-tuning is often undisclosed or opaque \citep{widder2023open}, while  the number of fine-tuned parameters is astronomical (on the order of $10^6$--$10^{11}$), rendering full re-estimation infeasible for downstream actors.

Researchers are divided on the recoverability of upstream fine-tuning structures.  \citep{Bommasani2021OnTO}. One camp advocates ``structural pessimism,'' emphasizing the near-impossibility of recovery and viewing fine-tuning as an ``alignment tax'' imposed on downstream actors \citep{ouyang2022training}. The opposing camp advances ``algorithmic optimism,'' appealing to statistical and engineering methods that, while have received significant academic attention,  have also been criticized as overly simplistic with limited effectiveness  \citep{mai2024finetuning, kumar2022fine}.\footnote{\citet{mai2024finetuning} first propose a calibration method that models fine-tuning as a simple transformation of the original parameters and recovers this transformation via estimation that relies the public availability of both data and models. The authors split a public dataset into two parts, train an open model on one part, and compare the results with those obtained from the other part.

Probing methods rest on the idea that alignment primarily modifies the output layer of the neural networks while leaving the intermediate layers intact. If so, one can prompt intermediate layers and conduct estimations about them. This approach works only if fine-tuning does not permeate all layers; empirical evidence suggests that its effect may be limited \citep{kumar2022fine}. } 

We join the structural pessimism camp. The downstream company in our model fixes the values of $\beta_{m}$, possibly based on state-of-the-art research. Yet even the best feasible guess may contain errors or fail to adjust to the continual updates by upstream foundation models. By contrast, dimensions  novel to the company's research design are not subject to fine-tuning, and their coefficients must be estimated by the company itself to predict consumer preferences. Prediction errors distort subsequent downstream decisions such as pricing, recommendation, and targeting.
\end{example}

\section{Analysis}\label{sec_analysis}

\subsection{Preliminaries}\label{sec_preliminary}
This section introduces the concepts and notation for the main analysis.

It is useful to express $Z=XJ$, where $J \in \mathbb{R}^{s \times r}$ denotes the mapping from the latent space to the population-coefficient space.  Under the assumption that $X$ has full column rank, $J$ is unique for any given $Z$. Partition $X$ as $[X_m, X_f]$, where $X_m \in \mathbb{R}^{T \times q}$ denotes the regressors associated with the misspecified coefficients, and $X_f \in \mathbb{R}^{T\times (s-q)}$ denotes those associated with the free coefficients. Partition $J$ accordingly as $J = [J_m^{\top}, J_f^{\top}]^{\top}$, where $J_m \in \mathbb{R}^{q \times r}$ and $J_f \in \mathbb{R}^{(s-q) \times r}$. The model can then be re-expressed as
\begin{equation}\label{eq:model2}
\tag{1'}
Y = \begin{bmatrix} X_m & X_f \end{bmatrix}
\left(
\begin{bmatrix} \beta_m \\ \beta_f \end{bmatrix}
+
\begin{bmatrix} J_m \\ J_f \end{bmatrix} \omega
\right)
+ \epsilon.
\end{equation}

For a square matrix $A$ of dimension $T$, define the Gram matrix for $A$ as
\[
G^A \coloneqq X'AX =
\begin{bmatrix}
X_m'AX_m & X_m'AX_f\\
X_m'AX_f & X_f'AX_f
\end{bmatrix}.
\]
The $(i,j)$ block of $G^A$, $G_{ij}^A$, is simply $X_i'AX_j$. In the case where $G_{ff}^A$ is invertible, let $G_{m\vert f}^A \in \mathbb{R}^{q \times q}$ denote its Schur complement in $G^A$, defined as
\[
G_{m\vert f}^A
\coloneqq
G_{mm}^A - G_{mf}^A (G_{ff}^A)^{-1} G_{fm}^A.
\]
We interpret Schur complement as residual information, by letting 
\[
P_{X_f}^A \coloneqq X_f (X_f' A X_f)^{-1} X_f' A,
\qquad
R_{X_f}^A \coloneqq Id - P_{X_f}^A
\]
denote the $A$-orthogonal projector onto the column space $\operatorname{col}(X_f)$ and the associated residual operator, respectively. Then
\[
G_{m\vert f}^A
=
X_m'AX_m
-
X_m'AX_f (X_f'AX_f)^{-1} X_f'AX_m
=
X_m' A R_{X_f}^A X_m=
(R_{X_f}^A X_m)' A (R_{X_f}^A X_m),
\]
where the last equality follows from the $A$-orthogonal decomposition: $X_m=P_{X_f}^A X_m+R_{X_f}^A X_m$. It follows that $G_{m\vert f}^A$ is the residual information in $X_m$ after $A$-projecting out the component in $\operatorname{col}(X_f)$.

Let $\succeq$ denote the Loewner order over matrices. For symmetric matrices $A$ and $B$ of the same size, $A \succeq B$ if and only if $x'Ax \geq x'Bx$ for all $T$-vector $x$, and  $A\succ B$ if $x'Ax>x'Bx$ for all $x\neq 0$. Using this order, we note the weak monotonicity of the Schur complement: $G^{A} \succ G^{B}$ implies $G^A_{m\vert f} \succeq G^{B}_{m\vert f}$.

\subsection{Distortion characterization}\label{sec_characterization}
Our first theorem full characterizes the prediction distortion.

\begin{theorem}\label{thm:bias}
$\Delta\hat{\omega}(Y_i)=-J_m'((G_{m\vert f}^{\Sigma^{-1}})^{-1}+J_mJ_m')^{-1}\delta.$ 
\end{theorem}

Prediction distortion arises as we map the vector $\delta$ of misspecified population coefficients to the latent space via $J_m'$. This process is mediated by a filter $((G_{m\vert f}^{\Sigma^{-1}})^{-1}+J_mJ_m')^{-1}$ that has two components. The term $G_{m\vert f}^{\Sigma^{-1}}$ represents the residual information in $X_m$ after $\Sigma^{-1}$-projecting out $X_f$, where $\Sigma^{-1}$ is the error precision matrix. The term $J_mJ_m'$ arises from the variance of $J\omega$, acting as a regularizer that attenuates the impact  of $G_{m\vert f}^{\Sigma^{-1}}$.

We develop the geometric intuition behind Theorem \ref{thm:bias} via a proof sketch. 

\vspace{-10pt}
\paragraph{Proof sketch.} Rewrite the first-stage estimation as
\[
\min_{\beta_f \in \mathbb{R}^{s-q}}\;(X_m\delta+X_f\beta_f)'M^{-1}(X_m\delta+X_f\beta_f).
\]
Viewing $X_m\delta$ as the dependent variable and $X_f$ as the regressor, this is standard GLS estimation with coefficient $-\beta_f$ and weighting matrix $M^{-1}$. The GLS estimator is
\[
\hat{\beta}_f
=
-(X_f'M^{-1}X_f)^{-1}X_f'M^{-1}X_m\delta.
\]
The resulting estimate of the outcome mean equals
\[
X\hat{\beta}
=
X_m\delta+X_f\hat{\beta}_f
=
R_{X_f}^{M^{-1}}X_m\delta,
\]
which is the residual after $M^{-1}$-projecting $X_m\delta$ onto $\operatorname{col}(X_f)$.

The BLP for $\omega_i$ given $\hat{\beta}$ and $Y_i$ equals
\[
\hat{\omega}(Y_i)=Z'M^{-1}(Y_i-X\hat{\beta}),
\]
whereas the undistorted BLP equals
\[
\omega^0(Y_i)=Z'M^{-1}Y_i.
\]
Taking the difference yields the BLP distortion
\[
\Delta\hat{\omega}(Y_i)=-Z'M^{-1}X\hat{\beta}.
\]
Substituting $X\hat{\beta}$ and $Z=X_m J_m + X_f J_f$ gives
\begin{align*}
\Delta\hat{\omega}(Y_i)
&=
-\left(J_m'X_m'+J_f'X_f'\right)
M^{-1}
R_{X_f}^{M^{-1}}X_m\delta \\
&=
- J_m'X_m'M^{-1}R_{X_f}^{M^{-1}}X_m\delta + 0 \\
&=
- J_m'G_{m\vert f}^{M^{-1}}\delta .
\end{align*}
The geometry is characterized by the $M^{-1}$-orthogonal projection of $X\hat{\beta}$ onto $\operatorname{col}(Z)$. Since $Z=XJ$, the projection decomposes into two: one onto $\operatorname{col}(X_f)$ and one onto $\operatorname{col}(X_m)$. The projection onto $\operatorname{col}(X_f)$ vanishes because $X\hat{\beta}$ is the residual after $M^{-1}$-projecting out $X_f$. The projection onto $\operatorname{col}(X_m)$ yields the quadratic form $X_m'M^{-1}R_{X_f}^{M^{-1}}X_m$, which is precisely the Schur complement $G_{m\vert f}^{M^{-1}}$.

It is crucial that both estimation and prediction use the same weighting matrix $M^{-1}$, so that the same projection operator applies across both stages.

It remains to isolate the roles of $X$, $J$, and $\Sigma$ in $G_{m\vert f}^{M^{-1}}$. This step relies on applications of matrix inversion identities and is deferred to Appendix~\ref{sec_proof}. \qed 

\vspace{5pt}
Notice two things. First, prediction distortion is constant across outcome realizations. Henceforth, we  denote it simply by $\Delta\hat{\omega}$. 

Second, prediction distortion vanishes as the residual information in $X_m$ becomes null.  The next proposition provides sufficient conditions for this to occur in the limit. 

\begin{proposition}\label{prop:vanish}
$\Delta\hat{\omega} \rightarrow 0$ as $G_{m\vert f}^{\Sigma^{-1}} \rightarrow 0$, which occurs in the limit as $\operatorname{col}(X_f)$ expands to contain $\operatorname{col}(X_m)$,  or as the error precision $\Sigma^{-1}$ vanishes. 
\end{proposition}

In light of Proposition \ref{prop:vanish}, we focus on situations in which residual information is non-null. We distinguish between multidimensional misspecification ($q>1$) and unidimensional misspecification ($q=1)$; the DM misspecifies multiple population coefficients in the first case and a single population coefficient in the second.  

The next proposition highlights the difficulty of deriving monotone comparative statics of prediction distortion  under multidimensional misspecification. By contrast, monotone comparative statics obtain under unidimensional misspecification.

\begin{proposition}\label{prop:monotone}
As $G_{m\vert f}^{\Sigma^{-1}}$ increases in the Loewner order, the following holds for each dimension $j=1,\cdots,r$ of $\Delta\hat{\omega}$:
\begin{enumerate}[(i)]
\item When $q>1$, the change is generally ambiguous.
\item When $q=1$, the magnitude weakly  increases, and the increase is strict if and only if the $j^{\text{th}}$ entry of $J_m$ (which has size $1\times r$ in this case) is nonzero.
\end{enumerate}
\end{proposition}




Below we identify two sources of non-monotonicity in the case of multidimensional misspecification, focusing on their geometric intuitions. Proposition \ref{prop:rater} of Section \ref{sec_application} supplements the discussion below  with analytical solutions in simple cases.

The first source of non-monotonicity is what we refer to as misalignment: typically, $\delta$ and the columns of $J_m$ are not collinear. For instance, when the latent space is unidimensional ($r=1$), $J_m$ is a $q$-vector. Prediction distortion in that case reduces to the inner product between $J_m$ and $\delta$, weighted by $((G_{m\vert f}^{\Sigma^{-1}})^{-1}+J_mJ_m')^{-1}$. Unless $\delta$ and $J_m$ are collinear, increasing $G_{m\vert f}^{\Sigma^{-1}}$ and, hence, the filter in the Loewner order can generate non-monotonic changes in this inner product. 

The second source of non-monotonicity arises from the filter itself: unless $G_{m\vert f}^{\Sigma^{-1}}$ and $J_mJ_m'$ commute, i.e., admit a common eigenbasis, changes in $G_{m\vert f}^{\Sigma^{-1}}$ rotate the eigenbasis of the filter, causing non-monotonic changes in the prediction distortion.

These complications disappear under unidimensional misspecification.  In that case, the filter is a scalar, and 
increasing $G_{m\vert f}^{\Sigma^{-1}}$ in the scalar sense increases 
the magnitude of the prediction distortion in each latent dimension. This result mirrors some of the conclusions in 
\cite{heidhues2025overconfidence}, which studies Gaussian learning with unidimensional misspecification 
in a different context. Our approach applies regardless of dimensionality.

Proposition \ref{prop:monotone} motivates two comparative statics exercises. In Section \ref{sec_cs1}, we examine the comparative statics of residual information, focusing on environmental shocks that yield a null impact on residual information and, consequently, on prediction distortion.  In Section \ref{sec_cs2}, we vary the alignment between $\delta$ and the left eigenvectors of $J_m$, drawing sharp distinctions between regimes in which the worst-case elasticity of prediction distortion is bounded and those in which it is unbounded.

\subsection{Comparative static: Residual information}\label{sec_cs1}
This section examines the comparative statics of prediction distortion with respect to residual information. The high-level question is how introducing free information to the system---in the sense defined below---influences prediction distortion.

\vspace{-8pt}

\paragraph{Informational treatments.} We consider four distinct informational treatments: adding free regressors, adding observations, adding novel sources, and enhancing error precision.

By adding free regressors,  we compare the original model \eqref{eq:model2}, re-expressed as
\[
Y=[X_m \; X_+ \mid X_f] \left( 
\begin{bmatrix} 
\beta_m \\ 
0 \\[-1.1em] 
\rule{1.2em}{0.4pt} \\[-0.4em] 
\beta_f 
\end{bmatrix} + 
\begin{bmatrix} 
J_m \\ 
0 \\[-1.1em] 
\rule{1.2em}{0.4pt} \\[-0.4em] 
J_f 
\end{bmatrix}\omega
\right) +\epsilon,
\]
with a new model
\begin{equation}\label{eq:column}
Y=[X_m  \mid X_+ \; X_f] \left( 
\begin{bmatrix} 
\beta_m
\\[-1.1em] 
\rule{1.2em}{0.4pt} \\[-0.4em] 
\beta_+ \\
\beta_f 
\end{bmatrix} + 
\begin{bmatrix} 
J_m \\[-1.1em] 
\rule{1.2em}{0.4pt} \\[-0.4em] 
J_+ \\
J_f 
\end{bmatrix}\omega 
\right) +\epsilon.
\end{equation}
We introduce a new block $X_+$ of regressors, whose population and latent  coefficients are null in the original model and become $\beta_+$ and $J_+\omega$, respectively, in the new model. Both  $\beta_+$ and $\beta_f$ are free coefficients estimated by the DM, while $\beta_m$ remains fixed at $\delta$.\footnote{An alternative formulation of adding regressors is to ``free up'' the population coefficients associated with $X_+$ from $0$ to $\beta_+$ while keeping $J_+$ fixed across treatments. It can be shown that the impact on residual information equals the effect described below plus an additional  term.} Therefore, the treatment adds no more misspecification but only regressors with unknown coefficients to the system.

By adding observations, we consider an augmented model with new rows
\begin{equation}\label{eq:row}
\begin{bmatrix} Y \\ Y_+ \end{bmatrix} = \begin{bmatrix} X \\ X_+ \end{bmatrix} (\beta + J\omega) + \begin{bmatrix} \epsilon \\ \epsilon_{+} \end{bmatrix},
\end{equation}
and an expanded error covariance matrix
\begin{equation}\label{eq:covariance}
\begin{bmatrix}
    \Sigma & \Sigma_{.+}\\
    \Sigma_{+.}&  \Sigma_{++}
\end{bmatrix}.
\end{equation}
The fixed and free coefficients: $\beta_m$ and $\beta_f$, remain unchanged relative to the original model. 

The third exercise: adding novel sources, is a hybrid of the first two. 
We introduce a new equation
\[
Y_+ = X_+(\beta_+ + J_+\omega) + \epsilon_+,
\]
and stack it with the original equation to obtain a block system 
\begin{equation}
\begin{bmatrix}
Y\\
Y_+
\end{bmatrix}
=
\begin{bmatrix}
X & 0 \\
0 & X_+
\end{bmatrix}
\left(
\begin{bmatrix}
\beta\\
\beta_+
\end{bmatrix}
+
\begin{bmatrix}
J\\
J_+
\end{bmatrix}
\omega
\right)
+
\begin{bmatrix}
\epsilon\\
\epsilon_+
\end{bmatrix},
\end{equation}
where the joint error covariance is as in  \eqref{eq:covariance}. Importantly, both $\beta_f$ in the original equation and $\beta_+$ in the new equation are free coefficients, while $\beta_m$ remains fixed at $\delta$. The new equation thus represents novel information sources whose population coefficients are unknown and must be estimated by the DM.

Enhancing error precision means strictly increasing the error precision matrix $\Sigma^{-1}$ in the Loewner order. 

All informational treatments have counterparts in our examples, as summarized in the next table. 

\begin{table}[htbp]
\centering
\caption{Informational treatments across applications.}
\label{tab:info_treatments}
\renewcommand{\arraystretch}{1.3}
\begin{tabular}{@{} 
p{0.22\textwidth} 
p{0.41\textwidth} 
p{0.34\textwidth} 
@{}}
\hline
Treatment & Employee rating & Consumer research \\ 
\hline
Add regressors 
& Rater treatment (de-bias training) 
& Enriched prompts \\ 

Add observations 
& Frequent evaluations; synthetic data 
& Higher query volume \\ 

Add novel sources 
& New raters (unknown biases) 
& In-house research integration \\ 

Enhance error precision 
& Precision-enhancing rater training 
& Upstream anti-hallucination measures \\ 
\hline
\end{tabular}
\end{table}

\vspace{-20pt}
\paragraph{Result.}   We say that an information treatment \emph{weakly (resp.\ strictly) increases residual information} 
if $G_{m\vert f}^{\Sigma^{-1}}$ weakly (resp.\ strictly) increases in the Loewner order after the treatment. 
The definition of reducing residual information is analogous.

The next theorem characterizes the effect of each informational treatment on residual information and, consequently, on prediction distortion.

\begin{theorem}\label{thm:residualinfo}
\begin{enumerate}[(i)]
    \item Adding free regressors weakly reduces residual information, with no change if and only if
    \[\left(R_{X_f}^{\Sigma^{-1}} X_m\right)' \Sigma^{-1} \left(R_{X_f}^{\Sigma^{-1}}X_+\right)=0.\]

    \item Adding observations weakly increases residual information, 
    with no change if
    \[
    X_+ = \Sigma_{+.}\Sigma^{-1} X.
    \]

    \item Adding novel sources weakly increases residual information, 
    with no change if and only if
    \[
    \Sigma_{+\cdot} \Sigma^{-1} 
    R_{X_f}^{\Sigma^{-1}} X_m
    \in \operatorname{col}(X_+).
    \]
    A sufficient conditions for this null condition is $\Sigma_{+.}=0$.
   
    \item Enhancing error precision weakly increases residual information.
\end{enumerate}

\vspace{8pt}
Prediction distortion is invariant under the null conditions in (i)--(iii). When $q=1$, the treatment in (i)  reduces the magnitude of each dimension of the prediction distortion, whereas those in (ii)-(iv) increase it. 
\end{theorem}

We sketch the proof of Theorem \ref{thm:residualinfo} in the remainder of this section. Applications of this theorem are deferred to Section \ref{sec_application}. 

\vspace{-10pt}
\paragraph{Proof sketch.} Part (i): The post-treatment linear projector is well-known to be 
\[
P_{[X_f, X_+]}^{\Sigma^{-1}}
=
P_{X_f}^{\Sigma^{-1}}
+
P_{r}^{\Sigma^{-1}},  
\quad \text{where} \quad
r\coloneqq R_{X_f}^{\Sigma^{-1}}X_+
\]
represents the additional explanatory power provided by the new regressor $X_+$ after projecting out its component in $\operatorname{col}(X_f)$. The null condition in Part (i)  states that $X_+$ does  not provide further explanation of $X_m$ beyond $X_f$. When this condition fails, the introduction of $X_+$ reduces the residual information in $X_m$. 

\vspace{8pt}
\noindent Part (ii): 
Applying the block matrix inversion identity to the error covariance matrix in \eqref{eq:covariance} gives 
\begin{align*}
\Sigma_{new}^{-1} &= \begin{bmatrix} \Sigma^{-1} + \Sigma^{-1}\Sigma_{\cdot+}S^{-1}\Sigma_{+\cdot}\Sigma^{-1} & -\Sigma^{-1}\Sigma_{\cdot+}S^{-1} \\ -\Sigma_{+\cdot}\Sigma^{-1}S^{-1} & S^{-1} \end{bmatrix} \\
&= \begin{bmatrix} \Sigma^{-1} & 0 \\ 0 & 0 \end{bmatrix} + \begin{bmatrix} \Sigma^{-1}\Sigma_{\cdot+} \\ -Id \end{bmatrix} S^{-1} \begin{bmatrix} \Sigma_{+\cdot}\Sigma^{-1} & -Id \end{bmatrix},
\end{align*}
where $S \coloneqq \Sigma_{++} - \Sigma_{+\cdot}\Sigma^{-1}\Sigma_{\cdot+}$ is the Schur complement. Inspecting the update term in the last line reveals that, if the original and new errors are correlated, then in $\Sigma_{new}^{-1}$, information leaks from the original diagonal block into the new via off-diagonal blocks. If we shut down this channel and set the off-diagonal blocks to zero, then the informativeness of the original errors would be artificially inflated. The $-Id$ term in the update corrects for this inflation and prevents double-counting. 
 
The post-treatment Gram matrix is given by 
\[G_{new}=\begin{bmatrix} X' & X_+' \end{bmatrix} \Sigma_{new}^{-1} \begin{bmatrix} X \\ X_+ \end{bmatrix}.\]
Substituting in the expression for $\Sigma_{new}^{-1}$ gives 
\begin{align*}
G_{new} 
= G_{old}+(\Sigma_{+\cdot}\Sigma^{-1}X - X_+)' S^{-1} (\Sigma_{+\cdot}\Sigma^{-1}X - X_+) \succeq G_{old},\end{align*}
and the inequality holds with equality if and only if \[X_+=\Sigma_{+\cdot}\Sigma^{-1} X.\] The term $\Sigma_{+\cdot}\Sigma^{-1}X$ represents the best linear predictor of $X_+$ given $X$.  If $X_+$ equals this term, then the informational correction discussed above is exactly offset by the linear relation between the original and new regressors. In that case, adding new observations does not change the Gram matrix. For all $X_+ \neq \Sigma_{+\cdot}\Sigma^{-1}X$, we have $G_{new}\succ G_{old}$. 

The conclusion then follows from the weak monotonicity of the Schur complement.  

\vspace{5pt}
\noindent Part (iii): This part builds upon the insights of Parts (i) and (ii). Details are in Appendix~\ref{sec_proof}. 

\vspace{5pt}
\noindent Part (iv): The Gram matrix $G^{\Sigma^{-1}}$ strictly increases with $\Sigma^{-1}$. Combine this with the weak monotonicity of the Schur complement. \qed 

\subsection{Comparative static: Alignment}\label{sec_cs2}
This section examines the comparative static of the prediction distortion as we rotate $\delta$, holding other model elements---including $X$, $\Sigma$, and $J_m$---fixed.

Given the nature of this exercise, it is w.l.o.g. to normalize $X=Id_s$ and $\Sigma=\sigma^2 Id_s$; see Online Appendix \ref{sec_standardization}. Post normalization,  $G^{\Sigma^{-1}}=X'\Sigma^{-1}X=\sigma^{-2} Id_s$, so $G_{m\vert f}^{\Sigma^{-1}}=\sigma^{-2} Id_q$ and commutes with $J_mJ_m'$. This rules out the eigenbasis-rotation complication discussed after the statement of Proposition \ref{prop:monotone}. In addition, we place $\delta \in S^{q-1}$ (the unit sphere in $\mathbb{R}^q$).

 The next lemma expresses the squared Euclidean norm of the prediction distortion in terms of the eigensystem of $J_mJ_m'$.

\begin{lemma}\label{lem:eigen}
Suppose $X=Id_s$ and $\Sigma=\sigma^2 Id_s$. Let 
\[
J_mJ_m' = U\Lambda U',
\qquad
U=[u_1, \cdots, u_q],
\quad
\Lambda=\operatorname{diag}(\lambda_1,\cdots,\lambda_q),
\]
where $U$ is orthonormal, and $\lambda_1 \ge \cdots \ge \lambda_q \ge 0$ are the eigenvalues of $J_mJ_m'$. Then  \[\|\Delta\hat{\omega}\|^2=\sum_{j=1}^q g(\lambda_j) b_j^2,\] where $b_j\coloneqq u_j'\delta$, and  \[
g(x)\coloneqq \dfrac{x}{(\sigma^2+x)^2}, \quad \mathbb{R}_+ \mapsto \mathbb{R}
\]
satisfies $g\ge0$, $\min g=g(0)=0$, and $\max g=g(\sigma^2)=(4\sigma^2)^{-1}$. 
\end{lemma}

We now develop our notation of alignment. Since $\delta$ and $u_j$, $j=1,\cdots, q$, all lie on the unit sphere,  $b_j=u_j'\delta$ represents the ``loading'' of $\delta$ onto $u_j$. Letting
\[\theta_j= \cos^{-1}\left(\frac{b_j}{\sqrt{\sum_{k\geq j} b_j^2}}\right) \in [0, \pi] \quad \text{for}\quad  j=1,\cdots, q-1,\] 
we can perform a spherical coordinate transformation: 
\[b_j=\left(\prod_{k<j} \sin(\theta_k)\right) \cos(\theta_j), \quad j=1,\cdots, q-1\quad \text{and} \quad b_q= \prod_{k=1}^{q-1} \sin(\theta_k).\]
Under the transformed coordinate system, $\theta_1$ captures the alignment between $\delta$ and the first principal direction $u_1$, $\theta_2$ captures the alignment between $\delta$ and the second principal direction $u_2$ after the component of $\delta$ in the first principal subspace $\operatorname{span}(u_1)$ is removed, and so forth. 

When $\theta_j=0$, $\delta$ is perfectly aligned with principal direction $u_j$. Increasing $\theta_j$ rotates $\delta$ away from $u_j$ and increases the total mass of loading onto directions $u_k$, $k>j$. Crucially, this change has no effect on the loading of $\delta$ onto directions $u_k$, $k<j$, or the angular relationship between $\delta$ and directions $u_k$, $k>j$.  This separability allows us to vary the alignment of $\delta$ and each $u_k$ in isolation, thereby facilitating comparative static analysis. 

\vspace{-10pt}
\paragraph{Result.} Write $\Delta\hat{\omega}(\theta)$ to make its dependence on 
$\theta\coloneqq[\theta_1 , \cdots, \theta_{q-1}]^{\top} \in [0,\pi]^{q-1}$ explicit. 
Consider the semi-elasticity of $\|\Delta\hat{\omega}(\theta)\|^2$ 
with respect to a marginal change in $\theta_j$:
\[
\mathcal{E}_j(\theta)
=
\frac{\partial \|\Delta\hat{\omega}(\theta)\|^2/\partial \theta_j}
{\|\Delta\hat{\omega}(\theta)\|^2}.
\]
The next theorem shows that the worst-case semi-elasticity may be bounded or unbounded, 
depending sharply on whether the minimum eigenvalue of $J_mJ_m'$ is zero or positive.\footnote{A natural extension is average-case analysis. Ongoing work by some coauthors models the misspecification vector and the latent-to-coefficient mapping as random. In high dimensions, prediction distortion is governed by the joint distribution of the regularizer’s eigenvalues and the misspecification vector, with asymptotic patterns emerging from random matrix theory. Details are left to future work.}

\begin{theorem}\label{thm:alignment}
Consider two cases:
\begin{enumerate}[(i)]
    \item Suppose $\lambda_q=0$. Then for each $j$ with $\lambda_j>0$,
    $
    \sup_{\theta} |\mathcal{E}_j(\theta)|=\infty.$
    \item Suppose $\lambda_q>0$. Then
    $
    \sup_{j, \theta} \left\vert\mathcal{E}_j(\theta)\right\vert<\infty.
$
\end{enumerate}
\end{theorem}

Theorem \ref{thm:alignment} highlights a critical  distinction between two regimes: $q>r$ and $q\le r$ (more or fewer misspecified coefficients than the latent space's dimensionality). When $q>r$, we necessarily have $\lambda_q=0$ and hence are in Case (i). Only when $q\le r$ is it possible to have $\lambda_q>0$, corresponding to Case (ii). We explore the implication of this dimensionality condition in Section \ref{sec_application}.

\vspace{-10pt}
    \paragraph{Proof sketch.} We only sketch the proof for Part (i) to build intuition. 
    
Fix any $j$ such that $\lambda_j>0$. For each $k=1,\cdots, q$, write $g_k$ for $g(\lambda_k)$. By Lemma \ref{lem:eigen}, we have $g_j>0$ and $g_q=0$. Below we construct a sequence of $b \in S^{q-1}$ (equivalently, $\theta \in [0,\pi]^{q-1}$) that sends $\mathcal{E}_j$ to infinity in the limit. 

First, let $b_k=0$ (equivalently, $\theta_k=\pi/2$) for all $k<j$. 
Intuitively, since adjusting $\theta_j$ does not affect the loadings of $\delta$ onto principal directions $k<j$, we may ignore these directions. Under this simplification, we can express 
\[
\mathcal{E}_j
=
\frac{2(R-1)\cot(\theta_j)}
{R+\cot^2(\theta_j)},
\quad
\text{where}\quad 
R \coloneqq 
\frac{\sum_{k>j} g_k b_k^2}{g_j \sum_{k>j} b_k^2}.
\]
As noted earlier, rotating $\delta$ away from principal direction $j$ increases its total loading onto directions $k>j$. 
The term $R$ captures the aggregate effect of directions $k>j$ relative to direction $j$. 

Now, let $b_j^2=\epsilon$, $b_k^2=0$ for $k=j+1,\ldots,q-1$, and $b_q^2=1-\epsilon$. 
That is, among directions $k \ge j$, we load primarily on direction $q$ and only slightly on direction $j$, with no mass elsewhere. 
Since $g_q=0$ while $g_j>0$, the numerator of $R$ equals zero while the denominator is positive. Letting $R=0$ in the expression for $\mathcal{E}_j$ yields
\[
|\mathcal{E}_j|=2|\tan(\theta_j)|,
\]
which grows to infinity as $\theta_j \to \pi/2$ (equivalently, $\epsilon \to 0$). 

Summarizing, we have
\[
b_k=\begin{cases}
0 & \text{ if } k<j \text{ or } j<k<q,\\
O(\epsilon) & \text{ if } k=j,\\
O(1-\epsilon) & \text{ if } k=q.
\end{cases} \quad \iff \quad 
\theta_k=\begin{cases}
\pi/2 &\text{ if } k<j \text{ or } j<k<q,\\
\approx \pi/2 & \text{ if } k=j,\\
\approx 0 &\text{ if } k=q.
\end{cases}\]
The worst case occurs when $\delta$ is initially aligned with directions associated with zero eigenvalues and is subsequently rotated toward a direction with a positive eigenvalue. The resulting elasticity in the prediction distortion's norm is unbounded,  as if an iceberg, previously submerged, suddenly breaks the surface of the water. \qed

\section{Applications}\label{sec_application}
This section examines the implications of our theorems for the examples outlined in Section~\ref{sec_application}.

\subsection{Employee rating with unconscious bias}\label{sec_application_discrimination}
In Example \ref{exm:labor}, the latent space of worker productivity is unidimensional ($r=1$).  We denote rater $j$'s precision by $\nu_j \coloneqq \sigma_j^{-2}$ and consider two scenarios labeled A and B, corresponding to unidimensional and multidimensional misspecification, respectively ($q=1$ or $q>1$). In each scenario, we establish a benchmark, followed by a sequence of interventions corresponding to the comparative statics in Theorems \ref{thm:residualinfo} and \ref{thm:alignment}.

\vspace{-10pt}
\paragraph{Scenario A.}  All cases in this scenario feature $q=1$. 
\vspace{-10pt}
\subparagraph{A0 (Benchmark).} There are two raters, one senior and one junior. The manager misspecifies the senior bias as $\delta \neq 0$ and must estimate the junior bias. 
The system is given by 
\[
X=\begin{bmatrix}
1 & 0\\
0 & 1
\end{bmatrix}, \quad 
\beta=\begin{bmatrix}
\delta\\
\beta_2
\end{bmatrix}, \quad 
J=\begin{bmatrix}
1\\
1
\end{bmatrix}, \quad 
\Sigma^{-1}=\begin{bmatrix}
\nu_1 & 0\\
0 & \nu_2
\end{bmatrix}.
\]

\vspace{-10pt}
\subparagraph{A1 (Add observation).}
To A0, we add a synthetic rater (rater 3) who shares the same bias as rater 2 but has an independent error. The resulting system is
\[
X=\begin{bmatrix}
1 & 0\\
0 & 1\\
0 & 1
\end{bmatrix}, \quad
\beta, J \text{ as in A0}, \quad
\Sigma^{-1}=\begin{bmatrix}
\nu_1 & 0 & 0\\
0 & \nu_2 & 0\\
0 & 0 & \nu_3
\end{bmatrix}.
\]
In reality, this change corresponds to more frequent evaluations by the same junior rater, enabled by high-frequency feedback systems involving regular ``check-ins'' and ``snapshots'' \citep{deloitte2015}. More recently, synthetic data are increasingly used to supplement employee performance analysis, especially when real evaluation data are limited or sensitive \citep{jayashankar2024advancing}.


\vspace{-10pt}
\subparagraph{A2 (Add regressor).}
To A1, we add a regressor representing rater treatment. The corresponding population coefficient \(\beta_3\) represents the treatment effect and must be estimated by the manager. We consider three variants depending on which raters are subject to the treatment:
\begin{figure}[h]
\centering
\begin{subfigure}[t]{0.31\textwidth}
\centering
\[
X=\begin{bmatrix}
1 & 0 & 0\\
0 & 1 & 1\\
0 & 1 & 0
\end{bmatrix}
\]
\caption{Applies to rater 2 only}
\label{int:addcolumn:a}
\end{subfigure}\hfill
\begin{subfigure}[t]{0.31\textwidth}
\centering
\[
X=\begin{bmatrix}
1 & 0 & 1\\
0 & 1 & 1\\
0 & 1 & 0
\end{bmatrix}
\]
\caption{Applies to raters 1 and 2}
\label{int:addcolumn:b}
\end{subfigure}\hfill
\begin{subfigure}[t]{0.31\textwidth}
\centering
\[
X=\begin{bmatrix}
1 & 0 & 1\\
0 & 1 & 1\\
0 & 0 & 1
\end{bmatrix}
\]
\caption{Applies to all raters}
\label{int:addcolumn:c}
\end{subfigure}
\end{figure}

\noindent Across all variants, we have 
\[\beta=\begin{bmatrix}
    \delta\\
    \beta_2\\
    \beta_3
\end{bmatrix}, \quad J=\begin{bmatrix}
     1\\
     1\\
     0
\end{bmatrix},\quad \Sigma^{-1}=\begin{bmatrix}
    \nu_1 & 0 & 0\\
    0 & \nu_2 & 0\\
    0 & 0 & \nu_3
\end{bmatrix}.\]

Our preferred interpretation of the treatment is de-biasing training aimed at reducing raters' prejudice and stereotypes. Traditional programs involve longitudinal interventions that emphasize continual learning, habit breaking, and mediated intergroup contact \citep{devine2012long}. Lighter-touch nudges have also been designed to achieve long-lasting effects at lower cost \citep{bohnet2016works}. More recently, \citet{kleinberg2018human} show that appropriately trained machine-learning algorithms help reduce bias in human decision-making.

\vspace{-10pt}
\subparagraph{A3 (Add novel source).}
To A0, we add an independent new rater (rater 3) whose bias $\beta_3$ is unknown and must be estimated by the manager. The resulting system is
\[
X=\begin{bmatrix}
1 & 0 & 0\\
0 & 1 & 0\\
0 & 1 & 1
\end{bmatrix}, \quad 
\beta=\begin{bmatrix}
\delta\\
\beta_2\\
\beta_3
\end{bmatrix}, \quad 
J=\begin{bmatrix}
1\\
1\\
1
\end{bmatrix}, \quad 
\Sigma^{-1}=\begin{bmatrix}
    \nu_1 & 0 & 0\\
    0 & \nu_2 & 0\\
    0 & 0 & \nu_3
\end{bmatrix}.
\]

In reality, this intervention corresponds to diversity efforts that increase minority representation in evaluation roles. A large literature surveyed by \citet{bertrand2017field} studies the anti-discrimination effects of increasing minority representation at upper echelons. The conventional wisdom is that introducing raters from traditionally underrepresented groups may bring fresh perspectives and dilute entrenched bias.

\vspace{-10pt}

\paragraph{Scenario B.} All cases in this scenario feature $q=2$. 
\vspace{-10pt}
\subparagraph{B0 (Benchmark)} Consider a variant of $A_0$ where the manager misspecifies both raters' biases as $\delta=[\delta_1, \delta_2]^{\top}$. The system is given by 
\[
X, J,  \Sigma^{-1} \text{ as in A0}, \quad  \beta=\delta
\]

\vspace{-15pt}
\subparagraph{B1 (Enhance error precision).} We increase both $\nu_1$ and $\nu_2$ relative to B0,     which in reality corresponds to precision-enhancing rater training. Many anti-discrimination programs emphasize that evaluative decisions should follow deliberate, structured procedures so that judgments are grounded in facts rather than instinct \citep{eberhardt2020biased}. See also \citet{kleinberg2018human} for evidence on ML-assisted decision-making to alleviate human errors.

\vspace{-10pt}
\subparagraph{B2 (Rotate misspecification vector).}
We rotate 
$\delta$ relative to B0 
while keeping its Euclidean norm constant. In practice, this corresponds to awareness programs that recalibrate the manager's beliefs about raters. In the case where rater $1$ represents the manager himself and rater $2$ an HR colleague, $\delta_1$ may be brought closer to zero through implicit bias training \citep{greenwald2020implicit}, whereas $\delta_2$ is addressed through enhanced accountability and transparency \citep{kalev2006best, raji2020closing}. To the extent that these programs differ in effectiveness and adoption speeds, they can rotate  $\delta$, in addition to reducing its norm.

\vspace{-10pt}

\paragraph{Result.} The next proposition specializes Theorem \ref{thm:bias} to the case of unidimensional latent space. For both scenarios A and B, we solve for the impact of each intervention relative to the benchmark.

\begin{proposition}\label{prop:rater}
    When \( r = 1 \),
\[
\Delta \hat{\omega}
=
-\frac{\langle J_m,\delta\rangle_{G_{m\vert f}^{\Sigma^{-1}}}}
{1+\|J_m\|_{G_{m\vert f}^{\Sigma^{-1}}}^2}.
\]
Therefore, in Scenario A, we have
\[
\Delta \hat{\omega}
=
-\frac{\gamma}{1+\gamma}\,\delta, 
\]
where
\[
\gamma =
\begin{cases}
\nu_1,
& \text{A0, A1, A2(a), A3}, \\[6pt]
\left(\nu_1^{-1}+\nu_2^{-1}+\nu_3^{-1}\right)^{-1},
& \text{A2(b)}, \\[10pt]
0,
& \text{A2(c)}.
\end{cases}
\]
In Scenario B, 
\[
\Delta \hat{\omega}
=
-\frac{\nu_1\delta_1  + \nu_2\delta_2 }
{1 + \nu_1 + \nu_2}.
\]
\end{proposition}

When $r=1$, $J_m$ is a $q$-vector. In the prediction distortion formula, the numerator is the inner product between $\delta$ and $J_m$, capturing their alignment. The denominator is the squared Mahalanobis norm of $J_m$, reflecting the regularizing effect of $J_mJ_m'$. The Schur complement $G_{m\vert f}^{\Sigma^{-1}}$ serves as the weighting matrix for both the numerator and the denominator.  

Relative to the general case, here the regularizer $J_mJ_m'$ is of rank one, so the filter $((G_{m\vert f}^{\Sigma^{-1}})^{-1}+J_mJ_m')^{-1}$ can be simplified using the Sherman-Morrison formula. Details are deferred to Appendix \ref{sec_proof}. 

\vspace{-10pt}
\paragraph{Implications: Scenario A.} Since $q=1$ in this scenario, the comparison of A0-A3 reduces that of the scalar-valued residual information, abbreviated as $\gamma$.

Inspecting  A0 reveals basic intuitions. If the senior rater is biased against minorities but the manager views him as neutral, then $\delta>0$. This negative unconscious bias lowers minority workers' compensation relative to majorities', if majority ratings are subject to a null or positive unconscious bias.  The distortion increases with rater precision, illustrating a potentially counterproductive effect of precision-enhancing training.

Next, compare A0 and A1-A3. Consistent with Theorem \ref{thm:residualinfo}, the distortion has a weakly larger under A1 than under A0. Under the present design, the difference is null because columns 1 and 2 of the regressor matrix remain linearly independent before and after the intervention. The same logic applies when we compare A1 and A2(a): since the treatment applies only to a junior rater, it does not provide explanatory power for senior status, hence columns 1 and 2 of the regressor matrix remain linearly independent.

Under A2(b), the treatment applies to the senior rater and one junior rater. Treatment status therefore partially explains seniority. Consequently, $\gamma$ decreases and the distortion is mitigated.

Finally, under A2(c), the treatment applies to all raters. Treatment status together with junior status fully determines seniority, so column 1 is spanned by columns 2 and 3, and the distortion vanishes.

This chain of comparisons raises a design question: rater training is resource-intensive and forms a billion-dollar industry involving academic researchers, consultants, technology vendors, and compliance units \citep{eberhardt2020biased,greenwald2020implicit}. If treatment must be limited to a subset of raters, which design is the most effective in closing the majority-minority compensation gap? The answer is to correlate treatment with rater seniority as under A2(b). By contrast, intervention A2(a), which targets junior raters only, is ineffective in reducing this  gap.\footnote{The majority-minority compensation gap clearly has implications for worker welfare. Fully characterizing the welfare consequences of the interventions requires that we specify the utility models of workers, firms, or even the social planner and is beyond the scope of the current paper.}

Intervention A3 doesn't change the distortion relative to A0. This result challenges the conventional wisdom that diversity efforts at the top can dilute entrenched bias. In the current setting, the damage caused by the misspecification of the senior bias cannot be not rectified by introducing new raters with unknown biases. Diversity initiatives at the top may operate through other channels, but not through this one.

\vspace{-10pt}
\paragraph{Implications: Scenario B.} This scenario features two misspecified raters. By Proposition \ref{prop:rater}, prediction distortion aggregates the precision-weighted misspecifications across them. 

Comparing B1 and B0 reveals that increasing precisions asymmetrically across raters has, in general, an ambiguous effect. For instance, increasing $\nu_1$ marginally while holding $\nu_2$ fixed exacerbates the distortion if and only if
\[
\operatorname{sgn}\left(\frac{\partial \Delta\hat{\omega}}{\partial \nu_1}\right)
=
\operatorname{sgn}(\Delta\hat{\omega})
\iff
\operatorname{sgn}(\nu_1\delta_1+\nu_2\delta_2)
=
\operatorname{sgn}((1+\nu_2)\delta_1-\nu_2\delta_2).
\]
Otherwise, the distortion is reduced. The practical implications of this finding are mixed: while precision-enhancing training may help, its success depends on a precise calibration to the underlying misspecification structure, which the manager is unaware of. We therefore remain cautious about the practicality of this approach. 

We finally compare B0 and B2. To apply our result, we set precisions equal across raters ($\nu_1=\nu_2=\nu$), and perform an eigenvalue decomposition:  
\[
J_mJ_m'=\begin{bmatrix}
    1 & 1 \\
    1 & 1
\end{bmatrix}
=U\Lambda U'
\quad \text{where}\quad
U=\frac{1}{\sqrt{2}}\begin{bmatrix}
    1 & 1\\
    1 & -1
\end{bmatrix},
\quad
\Lambda=\begin{bmatrix}
    2 & 0\\
    0 & 0
\end{bmatrix}.
\]
The smallest eigenvalue is zero because $q=2>r=1$, that is, the dimension of misspecification exceeds that of the latent space.

Part (i) of Theorem \ref{thm:alignment} examines, in this setting, the effect of rotating $\delta$, starting from a position   collinear with $u_2$ and then moving toward $u_1$, where
\[
u_1=\frac{1}{\sqrt{2}}\begin{bmatrix}
    1\\
    1
\end{bmatrix},
\qquad
u_2=\frac{1}{\sqrt{2}}\begin{bmatrix}
    1\\
    -1
\end{bmatrix}.
\]
The initial state corresponds to the situation in which raters 1 and 2 are biased in the opposite directions but with the same magnitude, yet the manager views both as neutral ($\delta_1=-\delta_2$). As the misspecifications cancel out, the distortion vanishes. A subsequent rotation of $\delta$ toward $u_1$, however small, leads to an unbounded elasticity in the squared norm of the distortion.

While the above discussion assumes unidimensional worker productivity, the conclusion remains valid even under multi-dimensional worker skills, provided there are more misspecified raters than the dimensionality of skills ($q>r$). If, instead, $q \le r$, and the minimum eigenvalue of $J_mJ_m'$ is positive, then the conclusion of the sensitivity analysis may be reversed.

\subsection{LLM-mediated consumer research}\label{sec_application_ai}
The techno-social consequences of adapting a small number of large-scale pre-trained models to downstream tasks have generated heated debate \citep{Bommasani2021OnTO}. In computer science, a central disagreement concerns whether downstream actors can undo the distortions introduced by  upstream fine-tuning. At present, the research community is divided into the ``structural pessimism'' camp and the ``algorithmic optimistic camp;'' see Section \ref{sec_application} for background review. 


We join the structural pessimism camp due to the intrinsic challenges mentioned earlier. In addition to the high-dimensionality and opacity of upstream fine-tuning, we note that state-of-the-art methods for recovering fine-tuning structure—such as calibration and probing—often rely on strong simplifying assumptions. While these assumptions provide useful starting points for research, they may be overly simplistic or adapt inadequately as upstream foundation models evolve. When downstream actors, lacking better alternatives, become overconfident in these methods and treat them as a ``holy grail,'' misspecification of the fine-tuning structure tends to follow.

 
Theorem \ref{thm:bias} quantifies the ``alignment tax'' paid by our downstream company due to misspecification of the fine-tuning structure. Theorem \ref{thm:residualinfo} cautions that recent developments, while considered as progresses in isolation, need not alleviate this tax. In LLM-mediated consumer research, adding observations corresponds to scaling query volume or using synthetic respondents \citep{brand2023using, ama2025llm}. Adding regressors corresponds to enriching inquiry design and introducing additional prompting dimensions, such as persona prompting and chain-of-thought prompting; the three subpanels (a)–(c) of Scenario A2 in Section \ref{sec_application_discrimination}
resemble the prompt-variation techniques studied in \cite{de2025ideation}. Adding novel sources corresponds to supplementing LLM-based surveys with independent in-house research \citep{millman2025firstparty}. Increasing error precision reflects anti-hallucination measures undertaken by upstream foundation model developers \citep{openai2024gpt4o}. 

Theorem \ref{thm:residualinfo} characterizes the conditions under which these developments have no effect on downstream distortion. When the effect is non-null, it depends sensitively on the exact misspecification structure. These findings provide new angles for assessing the aforementioned developments. For developments initiated by upstream actors—such as anti-hallucination measures—our results highlight their downstream and broader social consequences.

Theorem \ref{thm:alignment} examines the worst-case impact of upstream fine-tuning on downstream distortion, modeling adjustment in foundation models' alignment objectives as rotations of the misspecification vector $\delta$ relative to the eigenvectors of the regularizer $J_mJ_m'$. We show that the worst-case elasticity in downstream distortion is unbounded whenever $q>r$, that is, the dimensionality of fine-tuning exceeds that of the latent space. In consumer research, latent consumer preferences are typically estimated to lie between 32 and 521 \citep{wang2021exploring, awsPersonalizeHRNN}, i.e., on the order of $10^2$. By contrast, fine-tuning of modern LLMs modifies parameter spaces on the order of $10^9$ \citep{ouyang2022training}, and even parameter-efficient methods such as LoRA update millions of parameters \citep{hu2022lora}. The condition $q>r$ is therefore characteristic of consumer research in the AI era.

\section{Related literature}\label{sec_literature}

\paragraph{Misspecified Bayesian learning.} The study of misspecified Bayesian learning has a long tradition in statistical decision theory and, more recently, in economics; see \cite{bohren2024misspecified} for a survey. We broadly categorize the existing literature into passive learning models and active learning models.

In passive learning models, the DM excludes the true DGP from the support of his prior and updates beliefs based on repeated draws from the true DGP, without influencing it. The seminal work of \cite{berk1966limiting}, \cite{heidhues2025overconfidence} (HKS2025), and the current paper fall into this category. HKS2025 study Gaussian learning in which the DM misspecifies one dimension of the mean vector due to overconfidence about his own caliber. They characterize the KL-minimizing mean vector, which corresponds to the GLS estimator in our setting. Distortion in the KL minimizer informs how overconfidence about oneself induces discriminatory beliefs about in-groups versus out-groups. 

Our paper differs from HKS2025 in two main respects. First, our DM uses the GLS estimator to predict latent coefficients, a step absent from HKS2025. Thus, our analysis differs beyond the derivation of the GLS estimator.  Second, we focus on multidimensional misspecification, clarify the challenges they pose, and conduct comparative statics analysis tailored to this model feature.

In active learning models, the DM takes endogenous actions that influence the DGP. \cite{esponda2016berk} introduce Berk–Nash equilibrium as a steady-state notion, in which actions are optimal given beliefs, and beliefs concentrate on parameters that minimize the KL divergence from the DGP generated by equilibrium actions. More recent work studies the learning foundations of Berk–Nash equilibria.\footnote{These include, but are not limited to: single-agent models \citep{heidhues2018unrealistic, esponda2021asymptotic, fudenberg2021limit, frick2023belief}; multi-agent models \citep{murooka2021misspecified, ba2023multi, echenique2025implicit};  }  At present, models with infinitely many actions, states, and outcomes are scarce; exceptions such as \cite{heidhues2021convergence} and \cite{echenique2025implicit} adopt unidimensional parameterizations to gain insights into equilibrium stability. As the literature evolves, we expect increased attention to high-dimensional environments such as those studied in the current paper.


%
%


Models of misspecified Bayesian learning have been applied to other areas of economics such as discrimination and macrofinance. In the field of discrimination, beyond HKS2025, 
\cite{bohren2019dynamics} analyze a passive learning model in which a DM holds erroneous prior beliefs about minorities’ abilities and updates beliefs over time based on exogenous signals. The authors examine, both theoretically and experimentally, the evolution of beliefs and distinguish these from traditional taste-based or rational statistical discrimination.


\cite{echenique2025implicit} study an active learning model in which agents’ types consist of innate ability and effort productivity. A principal holds dogmatic, discriminatory beliefs about abilities and designs implicit incentives to induce costly effort. Effort generates noisy outcomes that update beliefs about productivity. The authors analyze the interaction between incentive provision and misspecified learning, deriving comparative statics of stable equilibria that inform anti-discrimination policies in education and labor markets.

In macrofinance, \cite{molavi2024model} (MOL2024) study a setting in which asset returns follow a high-dimensional hidden factor model, while market participants are restricted to using lower-dimensional models to explain observed returns. Both MOL2024 and the current paper analyze the prediction of latent variables under misspecified learning of the DGP, but differ in two key respects. First, MOL2024 consider a time-series environment with autocorrelated returns, whereas in our setting observations are independent conditional on the DGP. Second, the nature of misspecification differs: in MOL2024, misspecification arises from dimensionality constraints, with resulting distortions are characterized via principal component analysis; in our setting it takes the form of linear constraints on population coefficients. Consequently, both the theoretical results and the applications differ.

\vspace{-10pt}
\paragraph{Misspecification in econometrics/statistics.} The latent-variable data-generating process we consider has a long history in econometrics and statistics. It is known, modulo minor technical differences, as a linear mixed model, mixed-effects model, random coefficient model, random effects model, or hierarchical model. \cite{robinson1991blup} offers a classic review of the broad set of applications.  

To the best of our knowledge, the existing literature has not specifically derived our distortion of the latent-coefficient prediction. The closest line of work:  \cite{jiang2011best} and \cite{henderson2023improved}, studies prediction under misspecifications of similar natures to ours. The key difference lies in the loss function being minimized. In our work, the DM's worldview is distorted by the misspecified constraint; given that worldview, the constrained GLS and BLP are the best linear estimator and predictor. In stats, the analyst's objective---such as minimizing  unweighted prediction error---is not disciplined by our DM's worldview but is rather governed by  conventions. Importantly, such a choice violates our consistency condition on the weighting matrices  between estimation and prediction, yielding different results from ours. The ultimate goals also differ: statisticians evaluate the performance of their predictors using data, while we examine the distortion's comparative statics, relating it to the informational environment and alignment. 

A growing body of the literature investigates other forms of misspecification such as misspecified heteroskedasticity  \citep{chen2026empirical}, or focuses on different objects than BLP, such as confidence interval \citep{armstrong2022robust}.

Our analysis is conceptually related to hypothesis testing, which evaluates the impact of adding and removing constraints on the estimator (or hypotheses), absent latent coefficients (see  Chapter 2 of \citealp{jiang2021mixed}). A literature studies forecasting  with and without constrained specifications  (see \citealp{baltagi2013} for a comprehensive survey), under autoregressive DGPs and different loss criteria from ours. Its evaluation of forecasting performances is primarily empirical and numerical, differing from our comparative static approach.




\appendix

\section{Omitted proofs}\label{sec_proof}
\paragraph{Proof of Theorem \ref{thm:bias}.} In Section \ref{sec_characterization}, we already established that 
\[\Delta\hat{\omega}=-J_m' G_{m\vert f}^{M^{-1}} \delta.\]
It remains to show that 
\[G_{m\vert f}^{-1}=\left((G_{m\vert f}^{\Sigma^{-1}})^{-1}
+
J_mJ_m'\right)^{-1}.\]

Since $M^{-1}$ is invertible and $X$ has full column rank, 
$
G^{M^{-1}} = X'M^{-1}X$
is invertible. Consequently, the following identity of the Schur complement holds: 
\[
G_{m\vert f}^{M^{-1}}
=
\left(\big[(G^{M^{-1}})^{-1}\big]_{mm}\right)^{-1},
\]
where $[\cdot]_{mm}$ denotes the first $q\times q$ block of a matrix (the subscript ``$mm$'' stands for the ``misspecified'' block). 
Applying the Woodbury formula,
\[
M^{-1}
=
(\Sigma + ZZ')^{-1}
=
\Sigma^{-1}
-
\Sigma^{-1} Z (I + Z'\Sigma^{-1}Z)^{-1} Z'\Sigma^{-1}.
\]
Substituting into $G^{M^{-1}}$ yields
\begin{align*}
G^{M^{-1}}
&=
X'\Sigma^{-1}X
-
X'\Sigma^{-1}Z
(I+Z'\Sigma^{-1}Z)^{-1}
Z'\Sigma^{-1}X \\
&=
G^{\Sigma^{-1}}
-
G^{\Sigma^{-1}}J
(I+J'G^{\Sigma^{-1}}J)^{-1}
J'G^{\Sigma^{-1}}\\
&=\big((G^{\Sigma^{-1}})^{-1}+JJ'\big)^{-1},
\end{align*}
where the second equality uses the fact that $Z=XJ$, and the third equality uses the Woodbury formula again. Substituting into $G_{m\vert f}^{M^{-1}}$ gives
\[
G_{m\vert f}^{M^{-1}}=
\left(\big[(G^{\Sigma^{-1}})^{-1}\big]_{mm}
+
J_mJ_m'\right)^{-1}
=
\left((G_{m\vert f}^{\Sigma^{-1}})^{-1}
+
J_mJ_m'\right)^{-1},
\]
where the last step uses the identity of the Schur complement again.  \qed

\vspace{-10pt}
\paragraph{Proof of Theorem \ref{thm:residualinfo}.} We already established Parts (ii) and (iv) of the theorem in Section \ref{sec_cs1}. Here we establish Parts (i) and (iii). 

\vspace{5pt}

\noindent Part (i): We first elaborate on the proof presented in the main text. Let $P_{[X_f, X_+]}^{\Sigma^{-1}}$ denote the $\Sigma^{-1}$-orthogonal projector onto 
$\operatorname{col}(X_f) \oplus \operatorname{col}(X_+)$, where $\oplus$ denotes the direct sum. It is well known that 
\[P_{[X_f, X_+]}^{\Sigma^{-1}}=P_{X_f}^{\Sigma^{-1}}+P_r^{\Sigma^{-1}} \quad \text{where}\quad r=R_{X_f}^{\Sigma^{-1}} X_+.\]
Let $R_{[X_f, X_+]}^{\Sigma^{-1}}\coloneqq Id
-
P_{X_f}^{\Sigma^{-1}}
-
P_r^{\Sigma^{-1}}$ denote the corresponding residual operator. Following the introduction of $X_+$, the Schur complement becomes 
\[
G_{m\vert f+}^{\Sigma^{-1}} 
=
X_{m}' \Sigma^{-1}
R_{[X_f, X_+]}^{\Sigma^{-1}}
X_m.
\]
From the above, it follows that 
\begin{align*}
G_{m\vert f+}^{\Sigma^{-1}}
&=
X_m'\Sigma^{-1}\left(Id-P_{X_f}^{\Sigma^{-1}}\right)X_m
-
X_m'\Sigma^{-1}P_{r}^{\Sigma^{-1}}X_m \\
&=
G_{m\vert f}^{\Sigma^{-1}}
-
\left(P_r^{\Sigma^{-1}}X_m\right)'\Sigma^{-1}\left(P_r^{\Sigma^{-1}} X_m\right)\\
&\preceq 
G_{m\vert f}^{\Sigma^{-1}},
\end{align*}
where the last inequality holds with equality if and only if 
\[P_r^{\Sigma^{-1}}X_m =0\iff X_m' \Sigma^{-1} r=0 \iff \left(R_{X_f}^{\Sigma^{-1}}X_m\right)' \Sigma^{-1} r=0.\]
The first two conditions state that the columns of $X_m$ and $r$ are $\Sigma^{-1}$-orthogonal. The last condition follows from the $\Sigma^{-1}$-orthogonal decomposition:  $X_m=P_{X_f}^{\Sigma^{-1}}X_m+R_{X_f}^{\Sigma^{-1}}X_m$.

We next present an alternative proof that facilitates later analysis. It is well known that the Schur complement equals the minimized quadratic loss. That is, 
\begin{equation}\label{eq:schur1}
G_{m\vert f}^{\Sigma^{-1}} = \min_{B_f} \, (X_m + X_f B_f)' \Sigma^{-1} (X_m + X_f B_f), 
\end{equation}
and likewise
\begin{equation}\label{eq:schur2}
G_{m\vert f+}^{\Sigma^{-1}} = \min_{B_f, B_+} \, (X_m + X_f B_f + X_+ B_+)' \Sigma^{-1} (X_m + X_f B_f + X_+ B_+). 
\end{equation}
The relation $G_{m\vert f}^{\Sigma^{-1}} \succeq G_{m\vert f+}^{\Sigma^{-1}}$ is immediate.

To find conditions under which the equality holds, we solve the minimization in \eqref{eq:schur2} in two steps. First, for a fixed $e = X_m + X_f B_f$, we solve for the optimal $B_+$ as in the proof sketch of Theorem \ref{thm:bias}. The solution is 
\[ B_+^* = -(X_+' \Sigma^{-1} X_+)^{-1} X_+' \Sigma^{-1} e,\]
yielding an objective value of 
\[(R_{X_+}^{\Sigma^{-1}} e)' \Sigma^{-1} (R_{X_+}^{\Sigma^{-1}} e).\]
We then minimize this objective over $B_f$: 
\[ \min_{B_f}\; (R_{X_+}^{\Sigma^{-1}} e)' \Sigma^{-1} (R_{X_+}^{\Sigma^{-1}} e). \]
By contrast, the minimization in \eqref{eq:schur1} is 
\[\min_{B_f}\; e'\Sigma^{-1}e,\]
with solution $e^* = R_{X_f}^{\Sigma^{-1}} X_m$. The corresponding minimum value clearly upper bounds that of  \eqref{eq:schur2}, and the two are equal when 
\[R_{X_+}^{\Sigma^{-1}} e^* = e^*\iff X_+'\Sigma^{-1}e^* \iff \left(R_{X_f}^{\Sigma^{-1}}X_+\right)'\Sigma^{-1}e^*.\]
The first two conditions state that the columns of $X_+$ and $e^*=R_{X_f}^{\Sigma^{-1}}X_m$ are $\Sigma^{-1}$-orthogonal. The last follows from the $\Sigma^{-1}$-orthogonal decomposition: $X_+=P_{X_f}^{\Sigma^{-1}}X_+ +R_{X_f}^{\Sigma^{-1}}X_+$.

\vspace{5pt}

\noindent Part (iii): In the model with novel information sources, the Schur complement obtains by  solving 
\[ \min_{B_f, B_+} v' \Sigma_{new}^{-1} v \]
where 
\[ v \coloneqq \begin{bmatrix} X_m + X_fB_f \\ X_+B_+ \end{bmatrix}, \quad 
\Sigma_{new}^{-1} = \begin{bmatrix} \Sigma^{-1} & 0 \\ 0 & 0 \end{bmatrix} + \begin{bmatrix} \Sigma^{-1}\Sigma_{\cdot+} \\ -Id \end{bmatrix} S^{-1} \begin{bmatrix} \Sigma_{+\cdot}\Sigma^{-1} & -Id \end{bmatrix},
\]
Writing $X_m + X_fB_f=e$ as  in the proof of Part (i), we can expand the objective as
\[ e' \Sigma^{-1} e + (\Sigma_{+\cdot} \Sigma^{-1} e - X_+ B_+)' S^{-1} (\Sigma_{+\cdot} \Sigma^{-1} e - X_+ B_+). \]
For a fixed $e$, the optimal $B_+$ solves 
\[ \min_{B_+}\; (\Sigma_{+\cdot} \Sigma^{-1} e - X_+ B_+)' S^{-1} (\Sigma_{+\cdot} \Sigma^{-1} e - x_+ B_+) \]
and  equals
\[B_+^*=(X_+' S^{-1} X_+)^{-1} X_+' S^{-1} \Sigma_{+\cdot} \Sigma^{-1} e.\]
The objective given $e$ is therefore 
\[e'\Sigma^{-1}e+\left(R_{X_+}^{S^{-1}} (\Sigma_{+\cdot} \Sigma^{-1} e) \right)' S^{-1} \left( R_{X_+}^{S^{-1}} (\Sigma_{+\cdot} \Sigma^{-1} e) \right),\]
which is bounded below by $e'\Sigma^{-1}e$. 

By contrast, the original model without novel sources solves
\[ \min_{B_f} e' \Sigma^{-1} e, \]
with solution $e^* = R_{X_f}^{\Sigma^{-1}} X_m$. For the two problems to attain the same minimal value, we must have 
\[R_{X_+}^{S^{-1}}\Sigma_{+.}\Sigma^{-1}e^*=0 \iff  \Sigma_{+\cdot} \Sigma^{-1} e^*  \in \operatorname{col}(X_+).\]
A sufficient condition is simply $\Sigma_{+.}=0.$ \qed

\vspace{-10pt}




\paragraph{Proof of Proposition \ref{prop:rater}.}  

When \(r=1\), \(J_m\) is a \(q\)-vector, and \(J_mJ_m'\) has rank one. Applying the Sherman--Morrison formula to the filter gives
\[
\left((G_{m\mid f}^{\Sigma^{-1}})^{-1}+J_mJ_m'\right)^{-1}
=
G_{m\mid f}^{\Sigma^{-1}}
-
\frac{
G_{m\mid f}^{\Sigma^{-1}}J_mJ_m'G_{m\mid f}^{\Sigma^{-1}}
}{
1+J_m'G_{m\mid f}^{\Sigma^{-1}}J_m
}
\]
and consequently 
\begin{align*}
\Delta\hat{\omega}
=-J_m'\left((G_{m\mid f}^{\Sigma^{-1}})^{-1}+J_mJ_m'\right)^{-1}\delta
=
-\frac{
J_m'G_{m\mid f}^{\Sigma^{-1}}\delta
}{
1+J_m'G_{m\mid f}^{\Sigma^{-1}}J_m
}.
\end{align*}

The remainder of the proposition adapts this result to the specific settings studied in Section \ref{sec_application_discrimination}. Rather than presenting all algebraic derivations, we establish a slightly more general result that will be invoked in the online appendix: 


\begin{claim}
When $X=Id_s$, $J=\mathbf{1}_s$, and $\Sigma=\operatorname{diag}(\nu_1^{-1},\cdots, \nu_s^{-1})$, 
\[\Delta\hat{\omega}=-\bar\Delta \quad \text{where} \quad \bar\Delta\coloneqq \frac{\sum_{j=1}^q \delta_j \nu_j}{1+\sum_{j=1}^q \nu_j}.\]
\end{claim}

When sketching the proof of  Theorem \ref{thm:bias} in Section \ref{sec_characterization}, we already demonstrated that 
\[\hat{\beta}_f=(X_f'M^{-1}X_f)^{-1}X_f'M^{-1}X_m \delta.\]
Letting
\[X_m=\begin{bmatrix}
    Id_q\\
    0
\end{bmatrix}, \quad X_f=\begin{bmatrix}
    0 \\
    Id_{s-q}
\end{bmatrix} \]
in the expression for $\hat{\beta}_f$ simplifies it to 
\begin{align*}
\hat{\beta}_f=\left([M^{-1}]_{ff}\right)^{-1} [M^{-1}]_{fm}\delta.
\end{align*}
Moreover, since
\[M=\Sigma+(XJ) (XJ)'=\operatorname{diag}(\nu_1^{-1}, \cdots, \nu_s^{-1})+\mathbf{1}_s \mathbf{1}_s'\]
admits the form of a rank-one update to a diagonal matrix, we can calculate $M^{-1}$ using the Sherman-Morrison formula, which gives
\[M^{-1}=\operatorname{diag}(\nu)-\frac{1}{1+\mathbf{1}_s'\nu} \nu \nu' \quad \text{where}\quad \nu\coloneqq [\nu_1, \cdots, \nu_s]^{\top}.\]
Denoting $\nu_m\coloneqq [\nu_1,\cdots, \nu_q]^{\top}$ and $\nu_f=[\nu_{q+1},\cdots, \nu_s]^{\top}$, straightforward algebra yields 
\[[M^{-1}]_{fm}=-\frac{\nu_f \nu_m'}{1+\mathbf{1}_s' \nu}, \quad ([M^{-1}]_{ff})^{-1}=\operatorname{diag}(\nu_{f}^{-1})+\frac{1}{1+\mathbf{1}_q'\nu_m}\mathbf{1}_{s-q}\mathbf{1}_{s-q}',\]
and consequently 
\[([M^{-1}]_{ff})^{-1}[M^{-1}]_{fm}=-\frac{1}{1+\mathbf{1}_s'\nu}\left(1+\frac{\mathbf{1}_{s-q}'\nu_f}{1+\mathbf{1}_q' \nu_m}\right)\mathbf{1}_{s-q}\nu_m'=-\frac{1}{1+\mathbf{1}_q'\nu_m} \mathbf{1}_{s-q}\nu_m'.\]
Substituting these back gives
\[\tilde{\beta}_f=-\frac{1}{1+\mathbf{1}_q'\nu_m} \mathbf{1}_{s-q} \nu_m'\delta=-\bar\Delta\mathbf{1}_{s-q} \quad \text{and}\quad  \Delta\hat{\omega}=-\bar\Delta, \]
which establishes the claim. \qed 

\vspace{-10pt}

\paragraph{Proof Lemma \ref{lem:eigen}.} Under the assumptions that $X = Id_s$ and $\Sigma = \sigma^2 Id_s$, we have $G^{\Sigma^{-1}} = X'\Sigma^{-1}X = \sigma^{-2} Id_s$ and $G_{m\mid f}^{\Sigma^{-1}}=\sigma^{-2}Id_q$.  The prediction distortion is given by
\[
\Delta\hat{\omega} = -J_m' \left( \sigma^2 Id_q + J_m J_m' \right)^{-1} \delta,
\]
where $\sigma^2 Id_q$ and $J_mJ_m'$ commute. Since $J_m J_m'$ is symmetric and positive semi-definite, let $U \Lambda U'$ be its eigenvalue decomposition, where $U=[u_1, \dots, u_q]$ and $\Lambda = \operatorname{diag}(\lambda_1, \dots, \lambda_q)$. Then 
\[(\sigma^2 Id_q + J_m J_m')^{-1} = U \operatorname{diag}\left( \frac{1}{\sigma^2 + \lambda_k} \right) U',\]
and consequently 
\begin{align*}
\|\Delta\hat{\omega}\|^2 &= \delta' (\sigma^2 Id_q + J_m J_m')^{-1} J_m J_m' (\sigma^2 Id_q + J_m J_m')^{-1} \delta\\
&= \delta' U \operatorname{diag}\left( \frac{\lambda_k}{(\sigma^2 + \lambda_k)^2} \right) U' \delta \\
&= \sum_{k=1}^q g(\lambda_k) b_k^2,
\end{align*}
where $b_k = u_k' \delta$, and the function
$
g(\lambda) = \dfrac{\lambda}{(\sigma^2 + \lambda)^2}$, $\mathbb{R}_+\mapsto \mathbb{R}
$ 
satisfies $g \geq 0$,  $\min g=g(0)=0$, and $\max g= g(\sigma^2) = (4\sigma^2)^{-1}$. Henceforth, we shall write $g_k$ for $g(\lambda_k)$. \qed

\vspace{-10pt}
\paragraph{Proof of Theorem \ref{thm:alignment}.} The spherical coordinate transformation in Section \ref{sec_cs2} yields
\[b_j=\left(\prod_{k<j} \sin(\theta_k)\right) \cos(\theta_j), \quad j=1,\cdots, q-1 \quad \text{and} \quad b_q=\prod_{k=1}^{q-1} \sin(\theta_k). \]
Writing $b$ for $(b_k)_{k=1}^{q}$ and $\theta$ for $(\theta_j)_{j=1}^{q-1}$, straightforward algebra yields the Jacobian matrix: 
\[[\frac{\partial b}{\partial \theta}]_{kj}=\begin{cases}
0 & \text{ if } k<j,\\
-b_j\tan(\theta_j) & \text{ if } k=j,\\
b_k\cot(\theta_j) & \text{ if } k >j,
\end{cases}\]
and the semi-elasticity: 
\[
\mathcal{E}_j(\theta)
=\frac{\partial \|\Delta \hat{\omega}(\theta)\|^2/\partial \theta_j}{\|\Delta \hat{\omega}(\theta)\|^2}= \frac{\sum_{k=1}^q g_k \cdot 2 b_k \cdot (\partial b_k/\partial \theta_j)}
{\sum_{k=1}^q g_k b_k^2}.
\]

\noindent Part (i): We sketched the proof for this part in Section \ref{sec_cs2}. Here we provide the missing  details. 

Fix any $j$ with $\lambda_j>0$. We construct a sequence of $b$ (equivalently a sequence of $\theta$) that sends $|\mathcal{E}_j|$ to infinity in the limit.  

First, let $b_k=0$ for $k<j$. Then the denominator of $\mathcal{E}_j$ equals 
\[
0+g_j b_j^2 + \sum_{k>j} g_k b_k^2, 
\] 
whereas the numerator equals $1/2$ times 
\begin{align*}
0+ g_j b_j \frac{\partial b_j}{\partial \theta_j}+\sum_{k>j} g_k b_k \frac{\partial b_k}{\partial \theta_j}
&= -g_j b_j^2 \tan(\theta_j)+\left(\sum_{k>j} g_k b_k^2\right)\cot(\theta_j)\\
&=\left(\sum_{k>j} g_k b_k^2 - g_j \sum_{k>j} b_k^2\right)\cot(\theta_j),
\end{align*}
where the last equality uses the fact that $b_j^2/(\sum_{k>j} b_k^2)=\cot^2(\theta_j)$. 
Taking the ratio yields
\begin{align*}
\mathcal{E}_j=\dfrac{2\left(\sum_{k>j} g_k b_k^2 - g_j  \sum_{k>j} b_k^2\right)\cot(\theta_j)}{\sum_{k>j} g_k b_k^2+g_j b_j^2 }
=\frac{2(R-1)\cot(\theta_j)}{R+\cot^2(\theta_j)} \quad\text{where} \quad R= \dfrac{\sum_{k>j} g_k b_k^2}{g_j \sum_{k>j} b_k^2}.
\end{align*}

Next, let $b_j^2=\epsilon \in (0,1)$, $b_q^2=1-\epsilon$, and $b_k=0$ for $j<k<q$. Since $\lambda_j>0$ and $\lambda_q=0$, we have $g_j>0$ and $g_q=0$. Substituting yields $R=0$ and, in turn,
\[|\mathcal{E}_j|=\frac{2}{|\cot(\theta_j)|}=2\sqrt{\frac{1-\epsilon}{\epsilon}}.\]
Finally, send $\epsilon\rightarrow 0$ (equivalently $\theta_j \rightarrow \pi/2)$. 

\vspace{5pt}
\noindent Part (ii): Fix any $j=1,\cdots, q-1$. Since $\lambda_k>0$ for all $k=1,\cdots, q$ in this case,  we have $\min_{k=1,\cdots, q} g_k>0$. Therefore, the denominator of $\mathcal{E}_j$ can be bounded below as 
\[\sum_{k=1}^q g_k b_k^2 \geq \left(\min_{k} g_k\right) \cdot \left(\sum_{k=1}^q b_k^2\right)=\left(\min_{k} g_k\right) \cdot 1>0. \]
For the numerator, we can bounded it above as 
\begin{align*}
\left\vert\left(\sum_{k>j} g_k b_k^2 - g_j \sum_{k>j} b_k^2\right)\cot(\theta_j)\right\vert 
& \leq 2 \cdot \left(\max_{k} g_k\right) \cdot \left\vert \left(\sum_{k>j} b_k^2\right) \cdot \cot(\theta_j) \right\vert\\
&\leq 2 \cdot \left(\max g\right) \cdot \left\vert \left(\prod_{k<j}\sin^2(\theta_k)\right) \cdot \sin(\theta_j)\cos(\theta_j) \right\vert\\
& \leq (2\sigma^2)^{-1}.
\end{align*}
Combining these bounds yields 
\[|\mathcal{E}_j| \leq \frac{1}{\sigma^2 \min_{k} g_k}.  \]

\cleardoublepage

    \vspace*{16em}
    \begin{center}
        \Huge{
        Online Appendix for \\ ``Misspecified Model Estimation and Their Impact on Predictions''\\ by Junnan He, Lin Hu, Matthew Kovach and Anqi Li}
    \bigbreak
    \end{center}

\thispagestyle{empty}
\cleardoublepage

\appendix
\setcounter{section}{0}





\gdef\thesection{O.\arabic{section}}
\newtheorem{defnO}{Definition}
\renewcommand{\thedefnO}{O.\arabic{defnO}}
\newtheorem{exmO}{Example}
\renewcommand{\theexmO}{O.\arabic{exmO}}
\newtheorem{propO}{Proposition}
\renewcommand{\thepropO}{O.\arabic{propO}}

\section{Standardization}\label{sec_standardization}
Consider a general version of the statistical model: 
\begin{align}\label{eq:generalmodel}
Y_i = X \beta + Z \omega_i + \epsilon_i \\
\tag{true population coefficients} \text{where} \qquad  \beta^* \in \mathbb{R}^s \\
\tag{covariance matrices} \mathbb{V}[\omega_i] = \Omega, \quad \mathbb{V}[\epsilon_i] = \Sigma\\
\tag{linear constraints}\text{ and }\qquad R'\beta=\delta,
\end{align}
where the constraint matrix $R' \in \mathbb{R}^{q \times s}$ has full row rank. Compare this with the model in the main text, restated as
\begin{equation}
\tag{1} \tilde Y_i = \tilde X \tilde\beta + \tilde Z \tilde \omega_i + \tilde\epsilon_i \quad 
\text{where} \;  \tilde \beta^*=0, \; 
\mathbb{V}[\omega_i] = Id_r, \; \mathbb{V}[\epsilon_i] = \tilde \Sigma, \;
\text{ and } \; \beta_{1:q}=\tilde{\delta}. 
\end{equation}
We say that \eqref{eq:model} is a standardization of \eqref{eq:generalmodel} if it can be obtained via an invertible linear transformation of \eqref{eq:generalmodel}. Consequently, the two models define the same family of joint distributions over observables, and the results established for \eqref{eq:model} translate directly to \eqref{eq:generalmodel}.

The next proposition establishes two standardizations of the general model. The first is employed throughout the main text, and the second provides the refined structure needed in Section \ref{sec_cs2}.

\begin{propO}\label{lem_standardization1}
\begin{enumerate}[(i)]
\item 
There is a standardization of \eqref{eq:generalmodel} with $\tilde{\Sigma}=\Sigma$ and $\tilde{\delta}=\delta-R'\beta^*$. 

\item   There is a standardization of \eqref{eq:generalmodel} with $\tilde X = [Id_s,\;\; 0]^\top$ and $\tilde{\Sigma}=Id_T$. 
\end{enumerate}
\end{propO}

\begin{proof}
Part (i): The constraint matrix $R'$ has singular value decomposition (SVD)
\[
R' =  Q_R \begin{bmatrix}
    D_R & 0
\end{bmatrix} P_R',
\]
where $Q_R$ is orthonormal of dimension $q$, $P_R$ is orthonormal of dimension $s$,  and $D_R$ is diagonal of dimension $q$. Define 
\[
M_R:= 
\begin{bmatrix}
    Q_R D_R & 0 \\ 0 & Id_{s-q}
\end{bmatrix}P_R',
\]
and notice that $M_R$ is invertible with dimension $s$.

    Consider the following standardization:  
    \[
        \tilde Y_i := Y_i-\mathbb E[Y_i], \; 
        \tilde X := X M_R^{-1},\; 
        \tilde Z:= Z\Omega^{1/2},\;  \tilde \beta:=M_R(\beta-\beta^*), \;  \tilde{\omega}_i\coloneqq \Omega^{-1/2}\omega_i. 
    \]
    By construction, we have 
       \begin{align*}
        \tilde Y_i = Y_i-\mathbb E[Y_i] =X \beta +Z\omega_i + \epsilon_i -X\beta^* = \tilde X \tilde \beta  + \tilde Z\tilde \omega_i +\epsilon_i, 
    \end{align*}
   where $\tilde{\beta}$ has true value zero and $\tilde{\epsilon}_i$ has identity covariance: $\mathbb V[\tilde \omega_i]=Id_r$.  For the constraints, observe that 
    \[
    R' =\begin{bmatrix}
        Id_{q} & 0
    \end{bmatrix} M_R.
    \]
Since $\tilde \beta+M_R \beta^* =M_R\beta$, it follows that
\[
R' \beta =\delta \iff  
\begin{bmatrix}
        Id_{q} & 0
    \end{bmatrix} M_R \beta =\delta \iff  \beta_{1:q} = \delta-R' \beta^*.
\]
    This completes the proof of Part (i). 

\vspace{5pt}

\noindent Part (ii): In light of Part (i), it is w.l.o.g. to begin with a model 
with $\beta^*=0$, $\mathbb V[\omega_i]=Id_r,$ and $\beta_{1:q}=\delta$. 

Consider the matrix $\Sigma^{-1/2}X$ and its SVD: 
\[
\Sigma^{-1/2}X = Q  \begin{bmatrix}
    D  \\ 0
\end{bmatrix} P',
\]
where $Q$ is orthonormal of dimension $T$, $P$ is orthonormal of dimension $s$, and $D$ is diagonal with dimension $s.$  Define 
\[
    K:= \begin{bmatrix}
        Id_{q} & 0
    \end{bmatrix} P D^{-1} 
    =  Q_K \begin{bmatrix}
        D_K & 0
    \end{bmatrix} P_K',
\]
where $Q_K$ has  dimension $q$, $P_K$ has dimension $s$, and $D_K$ has  dimension $q$. Finally,   define
\[
    M_{obs} \coloneqq \begin{bmatrix}
    P_K'  & 0 \\
    0 & Id_{T-s}
\end{bmatrix}  Q', \quad 
    M_{par} := P_K' D P',
\]
and notice that both matrices are invertible. 

Consider the following standardization: 
\[
    \tilde Y_i \coloneqq M_{obs} \Sigma^{-1/2} Y_i, \; 
    \tilde X \coloneqq  M_{obs} \Sigma^{-1/2}X M_{par}^{-1}, \; 
    \tilde Z \coloneqq  M_{obs} \Sigma^{-1/2} Z, \;  
    \tilde \beta \coloneqq  M_{par}\beta. 
\]
It can be verified that $\tilde X= [Id_{s}, \; 0]^\top$, and that 
\begin{align*}
Y_i =& X \beta +Z\omega_i + \epsilon_i\\
& \iff \\
M_{obs} \Sigma^{-1/2} Y_i =& M_{obs} \Sigma^{-1/2}X M_{par}^{-1} M_{par}\beta +M_{obs} \Sigma^{-1/2}Z\omega_i + M_{obs} \Sigma^{-1/2}\epsilon_i\\
& \iff \\
    \tilde Y_i =& \tilde X \tilde \beta +\tilde Z\omega_i +\tilde \epsilon_i.
\end{align*}
Following this standardization, the constraints become 
\[
\begin{bmatrix}
    Id_{q} & 0
\end{bmatrix} \beta = \delta \iff \begin{bmatrix}
    Id_{q} & 0
\end{bmatrix} M_{par}^{-1}\tilde \beta = \delta 
\iff
K P_K \tilde \beta = \delta
\iff
    \begin{bmatrix}
    Id_{q} & 0
\end{bmatrix} 
 \tilde \beta =\tilde \delta\]
where $\tilde \delta:= D_K^{-1} Q_K' \delta$. Moreover, $\tilde{\epsilon}_i$ has identity covariance:  
    \[
\mathbb V[\tilde \epsilon_i] =M_{obs} \Sigma^{-1/2}\Sigma \Sigma^{-1/2} M_{obs}'  = Id_{T}.
\]
This completes the proof of Part (ii). \end{proof}

\section{Observation-specific regressors}\label{sec_varyingX}
 
In this appendix, we analyze a model variant: 
\begin{equation}\label{eq:observationspecificregressor}
    Y_i= \begin{bmatrix}
    X_1 & X_{2i}
\end{bmatrix} \left(\begin{bmatrix}
    \beta \\
    \gamma
\end{bmatrix} + \begin{bmatrix}
    J_1\\
    J_2
\end{bmatrix} \omega_i\right) + \epsilon_i,
\end{equation}
 where $X_{2i} $ is observation-specific. We continue to normalize the true population coefficients $[\beta^{\top},\; \gamma^{\top}]^{\top}$ to zero. The DM misspecifies $\beta_{1:q}$ as $\delta \neq 0$ while estimating both $\beta_{q+1:s}$ and $\gamma$ as free coefficients. 

The next proposition provides sufficient conditions under which the constrained GLS estimate of  $\gamma$ is asymptotically consistent. Consequently, when predicting $\omega_i$, the model collapses to the one analyzed in the main text: 
\begin{equation*}
    Y_i= \begin{bmatrix}
    X_1 & X_{2i}
\end{bmatrix} \left(\begin{bmatrix}
    \beta \\
    0
\end{bmatrix}+\begin{bmatrix}
    J_1\\
    0
\end{bmatrix}\omega_i  \right) + \epsilon_i =X_1 \left(\beta+J_1\omega_i\right) +\epsilon_i.
\end{equation*}

\begin{propO}
  In the setting of this appendix, suppose that  $J_2=0$ and that $X_{2i}$ has mean zero. 
Then the constrained GLS estimate of $\gamma$ under an infinite population equals its true value zero. 
\end{propO}

The assumption that $X_{2i}$ has mean zero is w.l.o.g. because one can always standardize the data of $X_{2i}$ across $i$'s. The condition $J_2=0$ isolates two sources of heterogeneity in the outcomes: those associated with fixed regressors $X_1$ and those associated with observation-specific regressors $X_{2i}$. The former  contribute to outcome heterogeneity through the latent coefficients $\omega_i$. The latter contribute  via themselves, with population coefficients $\gamma$.

\begin{proof}
Denote $X_i:=\begin{bmatrix}
    X_1 & X_{2i}
\end{bmatrix}$ and 
$v_i := X_i J \omega_i + \epsilon_i$. 
Rewrite the model as 
\[
Y_i = X_i  \begin{bmatrix}
    \beta \\
    \gamma
\end{bmatrix} + v_i, 
\]where $\mathbb V[v_i\mid X_i]=(X_iJ)(X_iJ)'+\Sigma\coloneqq M_i$. 

The constrained GLS estimation of $[\beta^{\top}, \gamma^{\top}]^{\top}$ under a finite sample of size $n$ solves 
\[
\min_{\beta, \gamma:\beta_{1:q}=\delta} \; \frac{1}{n}\sum_{i=1}^n\left(Y_i - X_i\begin{bmatrix}
    \beta \\
    \gamma
\end{bmatrix}\right)^\top M_i^{-1}\left(Y_i - X_i\begin{bmatrix}
    \beta \\
    \gamma
\end{bmatrix}\right).
\]
As $n \rightarrow \infty$, the objective tends to 
\begin{align*}
    &\lim_{n\to\infty} \frac{1}{n}\sum_{i=1}^n\left(-2Y_i^\top M_i^{-1} X_i\begin{bmatrix}
    \beta \\
    \gamma
\end{bmatrix} + \begin{bmatrix}
    \beta \\
    \gamma
\end{bmatrix}^\top X_i^\top M_i^{-1}X_i\begin{bmatrix}
    \beta \\
    \gamma
\end{bmatrix} \right) + \text{ constant }\\
= & \lim_n \begin{bmatrix}
    \beta \\
    \gamma
\end{bmatrix}^\top \left(\frac{1}{n}\sum_{i=1}^n X_i^\top M_i^{-1}X_i\right) \begin{bmatrix}
    \beta \\
    \gamma
\end{bmatrix} + \text{ constant},
\end{align*}
where the equality holds because outcomes have true mean zero. Under the additional assumptions stated  in the proposition, the above limit simplifies to 
\begin{align*}
   &    \lim_n \frac{1}{n}\sum_i^n \begin{bmatrix}
    X_1^\top \\ X_{2i}^\top 
\end{bmatrix} \left( X_i J J' X_i' +\Sigma\right)^{-1}  \begin{bmatrix}
    X_1 & X_{2i}
\end{bmatrix}  \\
= &  \lim_{n \rightarrow \infty} \frac{1}{n}\sum_i^n \begin{bmatrix}
    X_1^\top \\ X_{2i}^\top 
\end{bmatrix} \left( X_1J_1  J_1'X_1' +\Sigma\right)^{-1}  \begin{bmatrix}
    X_1 & X_{2i}
\end{bmatrix}   
\to \;    \begin{bmatrix}
    X_1^\top  M_1^{-1} X_1 & 0 \\ 0 & \mathbb E_i[X_{2i}^\top  M_1^{-1} X_{2i}]
\end{bmatrix},
\end{align*}
where $M_1:= (X_1J_1) (X_1J_1)'+\Sigma $.
Therefore, the population-level objective is 
\[
\begin{bmatrix}
    \beta \\
    \gamma
\end{bmatrix}^\top \begin{bmatrix}
    X_1^\top  M_1^{-1} X_1 & 0 \\ 0 & \mathbb E_i[X_{2i}^\top  M_1^{-1} X_{2i}]
\end{bmatrix} \begin{bmatrix}
    \beta \\
    \gamma
\end{bmatrix}=\beta^{\top} \left(X_1^\top  M_1^{-1} X_1\right)\beta+\gamma^{\top} \mathbb E_i[X_{2i}^\top  M_1^{-1} X_{2i}]\gamma
\]
and is minimized at $\gamma=0$. 
\end{proof}

\section{Joint estimation of $\beta$ and $\Sigma$}\label{sec_estimatecovariance}
In this appendix, we study a model variant where the DM estimates both the population coefficients $\beta$ and the error covariance $\Sigma$. The remaining model structure, including $X$, $J$, and $\Omega$, are known. For a given parameter pair $(\beta,\Sigma)$, the outcome distribution has mean $X\beta$ and covariance $M=\Sigma+(XJ)(XJ)'$. Let $(\beta^*,\Sigma^*)$ and $(\hat{\beta}, \hat{\Sigma})$ denote the true and estimated parameters, respectively.
Substituting them into the outcome distribution yields its first two moments under truth and estimation. 

We analyze two equivalent formulations, working directly with the outcome distribution. The first is a frequentist estimation solving two interdependent programs: first, $\hat{\beta}$ is the constrained GLS estimator given $\hat{M}$, solving
 \[\min_{\beta: \beta_{1:q}=\delta} (X\beta)'\hat{M}^{-1}(X\beta).\]
 Second, $\hat{M}$ is the sample second moment estimator for the outcome covariance given $X\hat{\beta}$, solving 
 \[\min_{M} \mathbb{E}[(Y-X\hat{\beta})(Y-X\hat{\beta})']. \]

The second formulation is Bayesian. Suppose that outcomes follow $\mathcal{N}(X\beta, M)$.  The DM chooses $(\beta, \Sigma)$ to minimize the KL divergence from the true distribution $\mathcal{N}(0, M^*)$, subject to the misspecification constraint: 
\[
\min_{(\beta,M):\\ \beta_{1:q}=\delta} \frac{1}{2}\left(\operatorname{tr}(M^{-1}M^*)-N+(X\beta)'M^{-1}(X\beta) +\log\frac{\operatorname{det} M}{\operatorname{det}M^*}\right).
\]


\begin{propO}\label{prop_jointestimate}
The following are true under joint estimation: 
\begin{enumerate}[(i)]
\item Compared to when $M^*$ is known, $\hat{\beta}$ remains unchanged, and $\hat{M}=M^*+(X\hat{\beta})(X\hat{\beta})'$, or equivalently  $\hat{\Sigma}=\Sigma^*+(X\hat{\beta})(X\hat{\beta})'.$
\item Let $G$ and $\hat{G}$ denote the Schur complements with weighting matrices $M^{*-1}$ and $\hat{M}^{-1}$, and let $D$ and $\hat{D}$ denote the corresponding prediction distortions. Then \[\hat{D}=\dfrac{D}{1+\delta'G\delta}.\]
\end{enumerate}
\end{propO}

On the one hand, joint estimation doesn't further distort the population-coefficient estimate relative to when the true error covariance is known. On the other hand, the estimated error  covariance is a rank-one update of the true covariance, implying that outcomes appear  ``noisier'' than they truly are in the Loewner order. Interestingly, such noises serve as a stabilizer that scales all dimensions of the prediction distortion by a uniform factor less than one.

\begin{proof}
Part (i): We only prove the result for the Bayesian program. Differentiating the KL divergence w.r.t. $M$ gives a first-order condition:  
\begin{align*}
  0=  \frac{\partial D_{KL}}{\partial M} =&
    \frac{ 1}{2} \left[ \frac{\partial}{\partial M}\text{tr}(M^{-1}M^*) +
    \frac{\partial}{\partial \Sigma}\text{tr} \left(M^{-1} (X\beta)(X\beta)'\right) 
    +M^{-1}\right] \\
    =&
    \frac{ 1}{2} \left( 
    -M^{-1}M^*M^{-1} 
    -M^{-1} (X\beta)(X\beta)'M^{-1}
    +M^{-1}\right).
\end{align*}
Thus for a given $\beta$,
\[
    M = M^*+(X\beta)(X\beta)' = \mathbb E [YY'] + (X\beta)(X\beta)' = \mathbb E [(Y-X\beta)(Y-X\beta)'],
\]
where the last equality uses the fact that $\mathbb{E}[Y]=0$. Letting $\hat{\beta}$ denote the GLS estimator in the baseline model and $\Delta\coloneqq X\hat{\beta}$, we can express $\hat{M}=M^*+\Delta\Delta'$. Since $\hat{M}$ is a rank-one update of $M^{*}$, 
\[\hat{M}^{-1}=M^{*-1}-\frac{M^{*-1} \Delta \Delta'M^{*-1}}{1+\Delta' M^{*-1}\Delta}\]
follows from the Sherman-Morrison formula. 

For a given $M$,  the first-order condition for GLS estimation is 
\[X'M^{-1}X\beta=0.\]
In the baseline model with $M=M^*$, this condition is $X'M^{*-1}\Delta=0$. Under joint estimation, substituting in $M=\hat{M}$ and $\beta=\hat{\beta}$ gives 
\[X'\hat{M}^{-1}X\hat{\beta}=X'\left(M^{*-1}-\frac{M^{*-1} \Delta \Delta'M^{*-1}}{1+\Delta' M^{*-1}\Delta}\right)\Delta=\frac{X'M^{*-1}\Delta}{1+\Delta' M^{*-1}\Delta}=0.\]
This establishes that $\hat{\beta}$ remains the GLS estimator of $\beta$ given $\hat{M}$. 

\vspace{5pt}
\noindent Part (ii): By the rank-one update formula  for the Schur complement, 
    \[G_{m\vert f}^{\hat{M}^{-1}}=G_{m\mid f}^{M^{*-1}}-\frac{\left(X_m'M^{*-1}R_{X_f}^{M^{*-1}}\Delta\right)\left(X_m'M^{*-1}R_{X_f}^{M^{*-1}}\Delta\right)'}{1+\Delta'M^{*-1}R_{X_f}^{M^{*-1}}\Delta}.\]
    To simplify the update term, recall that 
    \[\Delta=X\hat{\beta}=R_{X_f}^{M^{*-1}}X_m\delta.\]
  Since $R_{X_f}^{M^{*-1}}$ is idempotent,  
    \[R_{X_f}^{M^{*-1}}\Delta=(R_{X_f}^{M^{*-1}})^2 R_{X_f}^{M^{*-1}}X_m\delta=R_{X_f}^{M^{*-1}} R_{X_f}^{M^{*-1}}X_m\delta=\Delta.\]
   Substituting into the numerator and denominator of the update term gives
 \[X_m'M^{*-1}R_{X_f}^{M^{*-1}}\Delta=X_m'M^{*-1}\Delta=X_m'M^{*-1}R_f^{M^{*-1}}X_m\delta=G\delta
\]
    and \begin{align*}
    \Delta'M^{*-1}R_{X_f}^{M^{*-1}}\Delta&=\Delta'M^{*-1}\Delta\\
    &=\delta' X_m'R_{X_f}^{M^{*-1}}M^{*-1}R_{X_f}^{M^{*-1}}X_m\delta\\
    &= \delta' X_m'(R_{X_f}^{M^{*-1}})^2 M^{*-1}X_m\delta\\
    &= \delta' X_m'R_{X_f}^{M^{*-1}} M^{*-1}X_m\delta=\delta'G\delta,
    \end{align*}
    where the third equality uses the symmetry of $M^{*-1}$ and $R_{X_f}^{M^{*-1}}$, and the fourth equality uses again idempotency. It follows that 
    \[\hat{G}=G-\frac{(G\delta)(G\delta)'}{1+\delta'G\delta},\]
    and substituting into Theorem \ref{thm:bias} yields 
    \[\hat{D}=-J_m'\hat{G}\delta=-J_m'\left(G-\frac{(G\delta)(G\delta)'}{1+\delta'G\delta}\right)=-\frac{J_m' G\delta}{1+\delta'G\delta}=\frac{D}{1+\delta'G\delta}.\]
\end{proof}

\section{Misperceived media bias}\label{sec_media}
In this appendix, we apply the framework developed in the main text to the study of misperceived media bias. 
\vspace{-10pt}
\paragraph{Setup.}
In Example~\ref{exm:labor}, relabel raters \(j=1,\ldots,s\) as media outlets reporting on an evolving state over an infinite horizon.  Let $i \in \mathbb{N}$ index time and $\omega_i \in \mathbb{R}$ denote the period-$i$ state. States are i.i.d. standard normal. Outlet \(j\)'s report in period \(i\) is
\[
Y_{ij}=\beta_j+\omega_i+\epsilon_{ij},
\]
where \(\beta_j\in\mathbb{R}\) is outlet \(j\)'s partisan bias. The reporting error $\epsilon_{ij}$ is normal with mean zero and precision $\nu_j$, distributed independently across outlets and over time. 

Among all outlets, the first $q$ are traditional (e.g., CNN, Fox News), while the remainder are new (e.g., social media). The DM is a voter who observes reports from all media outlets. He believes that traditional outlets carry biases that he is familiar with, although such beliefs may be incorrect. Several cognitive factors may give rise to such perceptions. 
Suppose the DM is a Republican voter. He may perceive Fox News as neutral due to familiarity bias \citep{pennycook2021psychology}, or a bias blind spot that discounts the bias in his own views and those similar to them \citep{pronin2002bias}. Confirmation seeking and motivated reasoning may further lead him to discount the outlet's bias \citep{nyhan2020facts}. For the same reasons, he may perceive CNN as far left-wing and exhibit excessive hostility toward it \citep{hassell2020there}. These perceptions tend to strengthen with age, resist corrective information, and may even be transmitted across generations \citep{jennings2009politics}. We therefore model them as dogmatic but incorrect beliefs about the biases of traditional outlets.

By contrast, new media outlets are operated by humans or bots that the DM is less familiar with, so their biases must be estimated. Based on these, the DM casts a vote in period~$i$ that minimizes a quadratic loss $(a-\omega_i)^2$ in expectation. Under Gaussian assumptions, the optimal vote coincides with the BLP of $\omega_i$. Errors in the BLP translate into systematic differences in voting behavior across partisan voters.

\vspace{-10pt}
\paragraph{Result.} The system of media reports follows the same statistical model as in the employee rating example: 
\[Y_i=Id_s(\beta+\mathbf{1}_s \omega_i)+\epsilon_i.\]
In the proof of Proposition \ref{prop:rater}, we established that 
\[\hat{\beta}_f=-\bar\Delta \mathbf{1}_{s-q} \quad \text{ and } \quad \Delta\hat{\omega}=-\bar\Delta, \quad \text{where}\quad \bar\Delta\coloneqq \frac{\sum_{j=1}^q \nu_j \delta_j}{1+\sum_{j=1}^q \nu_j}.\]
The next proposition shows that $\bar\Delta$ also characterizes the distortion in the DM's objective welfare, defined as the difference in expected quadratic loss induced by the optimal behavior under the misspecified model and the correctly specified benchmark:
\[
\Delta W \coloneqq \mathbb{E}^*[(\omega_i-\omega^0(Y_i))^2]-\mathbb{E}^*[(\omega_i-\hat{\omega}(Y_i))^2]
\]
where the superscript ``$*$'' indicates that the expectation is taken with respect to the true distribution of $(\omega_i,Y_i)$. 

\begin{propO}\label{prop:media}
  In the setting of this appendix, 
    \[\hat{\beta}_f=-\bar\Delta\mathbf{1}_{s-q}, \quad \Delta\hat{\omega}=-\mathbf{1}_{s-q} \bar\Delta 
    \quad\text{and}\quad \Delta W=-\bar\Delta^2.\]
\end{propO}

\begin{proof}
Straightforward algebra shows that $\mathbb{E}^*[(\omega_i-\omega^0(Y_i))^2]=(1+\sum_{j=1}^s \nu_j)^{-1}$ and $\mathbb{E}^*[(\omega_i-\hat{\omega}(Y_i))^2]=(1+\sum_{j=1}^s \nu_j)^{-1}+\bar\Delta^2$. 
\end{proof}

The remainder of this appendix refers to $\bar\Delta$ as the distortion measure and examines the  implications of Proposition \ref{prop:media}. 

\vspace{-10pt}
\paragraph{Hostile media effect and false polarization.}
Our Republican DM underestimates the conservative bias of Fox News and overestimates the liberal bias of CNN. Under the normalization of true biases to zero, we have $\delta_j<0$ for both outlets. The distortion of the DM's worldview therefore manifests as a liberal media bias of magnitude $|\bar\Delta|$. Proposition \ref{prop:media} establishes a mechanism through which this distortion bias the DM's estimates $\hat{\beta}_f$ of new media outlets.  A conservative media bias among Democratic voters obtains analogously.

This finding relates to the hostile media effect, referring to opposing partisans' tendency to perceive news coverage as more biased against their side than it actually is. Ample empirical evidence in the political science, communication, and psychology documents the existence of this phenomenon in traditional media \citep{feldman2014hostile},\footnote{In a classic experiment conducted by \cite{arpan2003experimental}, student subjects read a balanced story about their hometown college football team published in one of three newspapers: a hometown paper, a rival-town paper, or a neutral paper outside the state. Subjects perceived the coverage as favoring the rival university even when it was believed to come from the neutral paper. More recently, \cite{hassell2020there} present journalists with balanced news stories that differ only in their ideological content. Democratic journalists exhibit no gatekeeping bias when selecting stories to cover---a finding that challenges the widely perceived ``liberal media bias'' among Republicans.} while research on new media remains sparse. Our model shows how hostile  perceptions of traditional outlets can induce similar perceptions of new outlets.


False polarization---the difference between the perceived distance between  two groups and the actual distance between them---is present in the American public \citep{levendusky2016mis}.\footnote{The authors asked respondents in a nationally representative sample to report their own positions on an issue scale and to place a typical Democratic and Republican voter on the same scale. They found that Americans significantly overestimate the extent of partisan disagreement.} Because individuals can now express their opinions online at almost no cost, our model predicts false polarization in the digital news sphere. Consistent with the literature's common attribution of false polarization to partisan media exposure \citep{levendusky2016does}, our model reveals a causal link between the two phenomena.

\vspace{-10pt}
\paragraph{Voting behavior.}
The existing literature is divided on the impact of (mis)perceived media bias on political behavior. While some authors, such as \cite{levendusky2016does}, document a moderation effect arising from false polarization, our result suggests a more polarized pattern, as reflected in the BLP distortion.

\vspace{-10pt}
 \paragraph{Fact-checking.} Proposition \ref{prop:media} establishes that features of new media outlets---their quantities, precisions, biases---are irrelevant to the distortion measure $\bar\Delta$. As we increase the precisions of these outlets, the DM's expected utility under the correctly specified benchmark increases unambiguously. Since the welfare loss $\bar\Delta^2$ due to misspecification is independent of these precisions, the DM's objective welfare under the misspecified model also increases. The same cannot be said of increasing the precisions of traditional media, which enter directly into the distortion measure.

This finding suggests that policy interventions targeting misinformation and fake news by improving source accuracy (e.g., through fact-checking) may entail unintended welfare consequences, especially when they operate on traditional sources about which people hold dogmatic misperceptions. By contrast, when applied to new sources, fact-checking unambiguously improves voter welfare. Given that professional fact-checking is resource intensive and does not scale easily, the allocation of resources should prioritize new sources over traditional ones. To our knowledge, this finding is new to the literature on (political) misinformation (see \citealt{nyhan2020facts} for a survey).

\vspace{-10pt}

\paragraph{Misperception recalibration.} A large empirical literature studies the treatment effect of misperception recalibration \citep{bursztyn2022misperceptions}, which in our context means recalibrating the DM's perceptions about traditional outlets to their true biases via, for example, the provision of corrective information or cross-cutting exposure. The effectiveness of such recalibration remains the subject of active research: early research pointed to the possibility of a ``backfire effect,'' whereas more recent work has shifted attention to alternative mechanisms \citep{nyhan2020facts}.

Our analysis reveals a new concern. If misperception recalibration has heterogeneous effectiveness across outlets, then it changes the orientation of the misspecification vector relative to the eigenbasis of the regularizer. In our setting, the DM misspecifies the biases of both Fox News and CNN while the underlying state space is unidimensional, hence $q=2>r=1$. By Theorem \ref{thm:alignment}, the distortion measure $\bar\Delta$ can exhibit unbounded sensitivity to recalibration efforts. In more general environments with a multidimensional issue space ($r>1$), the regime $q>r$ remains empirically relevant given the proliferation of traditional media outlets.

\newpage

\begin{spacing}{.8}
\bibliographystyle{aer} 
\bibliography{misspecifiednews.bib}
\end{spacing}
 \end{document}